\documentclass[12pt,a4paper]{article}
\usepackage{graphicx,psfrag,epsf,color}
\usepackage[dvipsnames,table]{xcolor}
    \definecolor{Lightgray}{gray}{0.95}
\usepackage{multirow}
\usepackage{amsmath,amssymb,amsfonts,bm}
\usepackage[
    bookmarksnumbered=true,
    urlcolor=blue,
    linkbordercolor=red,
    citebordercolor=green,
    bookmarksopen=true
    ]{hyperref}
\usepackage{appendix}
\usepackage{multirow}
\usepackage{geometry}
\usepackage{float}
\usepackage{mathrsfs}  
\usepackage{authblk}
\usepackage{array}
\usepackage{hhline}
\usepackage{cleveref}

\usepackage{cite}
\usepackage{slashed}
\usepackage{graphicx}
\usepackage{rotating}
\usepackage{slashed, cancel}
\usepackage{diagbox}

\usepackage{tabularx}
\newcolumntype{C}{>{\centering\arraybackslash}X}

\renewcommand{\arraystretch}{1.25}

\newcommand{\tr}[1]{\langle #1 \rangle}
\newcommand{\com}[2]{[#1,#2]}
\newcommand{\anticom}[2]{\{#1,#2\}}

\newcommand{\T}{\bm{T}}
\renewcommand{\t}{\bm{t}}
\newcommand{\Id}{\bm{1}}
\newcommand{\F}{\bm{F}}
\newcommand{\D}{\bm{D}}
\newcommand{\DELTA}{\bm{\Delta}}
\newcommand{\NABLA}{\bm{\nabla}}

\renewcommand{\S}{\bm{S}}

\newcommand{\Aj}{\bm{A}^{\!(j)}}
\renewcommand{\O}{\bm{O}}
\newcommand{\Oj}{\bm{O}^{(j)}}
\renewcommand{\H}{\bm{\mathcal{H}}}
\newcommand{\W}{\bm{\mathcal{W}}}
\newcommand{\VG}{\bm{V}^G}
\newcommand{\GammaH}{\bm{\Gamma}^H}
\newcommand{\GammaS}{\bm{\Gamma}^S}
\newcommand{\Gammac}{\bm{\Gamma}^c}
\newcommand{\GammaC}{\bm{\Gamma}^C}

\newcommand{\GammaBar}{\overline{\bm{\Gamma}}}
\newcommand{\Rc}{\bm{R}^c}

\newcommand{\Vc}{\bm{V}^c}

\numberwithin{equation}{section}

\renewcommand{\title}{Glauber Phases in Non-Global LHC Observables: \texorpdfstring{\\[1ex]}{}Resummation for Gluon-Initiated Processes}

\begin{document}
\allowdisplaybreaks

\begin{titlepage}

\begin{flushright}
{\small
MITP-23-064\\
November 30, 2023
}
\end{flushright}

\vskip0.8cm
\pdfbookmark[0]{\title}{title}
\begin{center}
{\Large \bf\boldmath \title}
\end{center}

\vspace{0.5cm}
\begin{center}
    \textsc{Philipp Böer,$^a$ Patrick Hager,$^a$ Matthias Neubert,$^{a,b}$ \\ Michel Stillger$^a$ and Xiaofeng Xu$^a$} \\[6mm]
    
    \textsl{${}^a$PRISMA$^+$ Cluster of Excellence \& Mainz Institute for Theoretical Physics\\
    Johannes Gutenberg University, 55128 Mainz, Germany\\[0.3cm]
    ${}^b$Department of Physics \& LEPP, Cornell University, Ithaca, NY 14853, U.S.A.}
\end{center}

\vspace{0.6cm}
\pdfbookmark[1]{Abstract}{abstract}
\begin{abstract}
\vskip0.2cm\noindent
The resummation of the ``Glauber series'' in non-global LHC observables is extended to processes with gluons in the initial state. 
This series simultaneously incorporates large double-logarithmic corrections, the so-called ``super-leading logarithms'', together with higher-order exchanges of pairs of Glauber gluons associated with the large numerical factor $(i\pi)^2$.
On a technical level, the main part of this work is devoted to the systematic reduction of the appearing color traces and construction of basis structures, which consist of thirteen elements for $gg$ and eleven elements for $qg$ scattering. 
Numerical estimates for wide-angle gap-between-jet cross sections at the parton level show that, in particular for $gg$ scattering at relatively small vetoes $Q_0$, the contribution involving four Glauber exchanges gives a sizeable correction and should not be neglected. 
\end{abstract}

\vfill\noindent\rule{0.4\columnwidth}{0.4pt}\\
\hspace*{2ex} {\small \textit{E-mail:} \href{mailto:pboeer@uni-mainz.de}{pboeer@uni-mainz.de}, \href{mailto:pahager@uni-mainz.de}{pahager@uni-mainz.de}, \href{mailto:matthias.neubert@uni-mainz.de}{matthias.neubert@uni-mainz.de}, \\
\hspace*{2ex} \phantom{E-mail: } \href{mailto:m.stillger@uni-mainz.de}{m.stillger@uni-mainz.de}, \href{mailto:xiaxu@uni-mainz.de}{xiaxu@uni-mainz.de}}

\end{titlepage}

{\hypersetup{hidelinks}
\pdfbookmark[1]{Contents}{ToC}
\tableofcontents}
\clearpage

%%%%%%%%%%%%%%%%%%%%%%%%%%%%%%%%%%%%%%%%%%%%%%%%%%%%%%%%%%%%%%%%%%
\section{Introduction}
\label{sec_intro}

The Standard Model of Particle Physics (SM) stands resolute, providing an accurate description of Nature at the smallest currently observable distance scales.
Nonetheless, both empirical observations and theoretical considerations suggest the existence of New Physics beyond the SM.
Moreover, it becomes increasingly evident that the theoretical precision must keep pace with the experimental advancements to fully harness the discovery potential of the LHC.
Evidence of deviations from the SM may already be present in the available datasets, clouded by theoretical uncertainties of the predictions.
Therefore, a meticulous control of the theoretical accuracy is paramount.

Jet observables play a pivotal role at high-energy colliders.
They directly probe the underlying hard-scattering process, thereby providing a unique vantage point to study the strong interactions at the shortest distances, where the scattering can be computed perturbatively.
The distinction of highly collimated energetic particles constituting the jet from the ``soft'' out-of-jet radiation by imposing veto criteria in certain parts of the phase space leads to \emph{non-global} observables~\cite{Sterman:1977wj}.
Such vetoes introduce additional scales, which in turn cause large logarithmic corrections.
For example, in gap-between-jets cross sections, the soft energy is restricted outside the jet by introducing a low jet-veto scale $Q_0$ much smaller than the large scale $Q \sim \sqrt{\hat{s}}$ of the process determined by the partonic center-of-mass energy. 
Consequently, large logarithms in the scale ratio $L=\ln(Q/Q_0)$ occurring in perturbative calculations of such observables must be controlled beyond the fixed-order expansion.
For example, so-called non-global logarithms arise from soft-gluon radiation off secondary emissions inside the jets~\cite{Dasgupta:2001sh}.
However, an all-order resummation of these logarithms is highly non-trivial due the intricate pattern of these corrections and the complexity of the color-algebra involved. 

Moreover, an additional subtlety manifests at hadron colliders.
Due to the exchange of Glauber gluons between the initial-state partons, color coherence breaks down leading to a non-cancellation of collinear singularities~\cite{Catani:2011st,Forshaw:2012bi,Schwartz:2017nmr}.
This gives rise to a series of double-logarithmic corrections starting at four-loop order, the so-called ``super-leading logarithms'' (SLLs)~\cite{Forshaw:2006fk,Forshaw:2008cq,Keates:2009dn}.
The all-order resummation of this double-logarithmic series has been achieved recently for generic \mbox{$2\to M$} partonic scattering processes in \cite{Becher:2021zkk,Becher:2023mtx}, by exploiting the renormalization-group (RG) equations of an effective field theory (EFT) description for jet processes~\cite{Becher:2015hka,Becher:2016mmh,Balsiger:2018ezi} based on soft-collinear effective theory (SCET)~\cite{Bauer:2000yr,Bauer:2001yt,Beneke:2002ph,Beneke:2002ni}.
This series is of the form
\begin{equation} \label{eq_SLLgeneric}
   \sigma^{\rm SLL} \sim \frac{\alpha_s\,L}{\pi\,N_c} \left( \frac{N_c\,\alpha_s}{\pi}\,\pi^2 \right)
    \sum_{n=0}^\infty\,c_{1,n} \left( \frac{N_c\,\alpha_s}{\pi}\,L^2 \right)^{n+1} \equiv \, \frac{\alpha_s\,L}{\pi\,N_c} \, w_\pi
    \sum_{n=0}^\infty\,c_{1,n} w^{n+1} \,,
\end{equation}
with the two parameters $w = \frac{N_c\,\alpha_s}{\pi}\,L^2$ and $w_\pi = \frac{N_c\,\alpha_s}{\pi}\,\pi^2$.
The SLLs require two Glauber phases, which are manifest in the prefactor $w_\pi$ in~\eqref{eq_SLLgeneric}, reflecting the square of the imaginary part of the large logarithm $\ln[(-Q^2-i0)/Q_0^2] = 2L-i\pi$.
The prefactor in~\eqref{eq_SLLgeneric} indicates that this series is subleading in the large-$N_c$ expansion. 
An analytic understanding of such effects will be beneficial in the ongoing effort to improve parton showers beyond the large-$N_c$ limit, see e.g.~\cite{Nagy:2019pjp,DeAngelis:2020rvq,Hoche:2020pxj,Hamilton:2020rcu,Becher:2023vrh}

The parameters $w,w_\pi$ are of $\mathcal{O}(1)$ for typical scales at the LHC, thus motivating an investigation of the higher-order corrections in both parameters $w$ and $w_\pi$. 
Facilitated by the fact that the color algebra simplifies drastically for initial-state partons in the fundamental representation, the resummation including arbitrarily many Glauber phases in addition to the SLLs has been achieved in~\cite{Boer:2023jsy} for all (anti-)quark-initiated processes. 
The generalized ``Glauber series'' is an alternating series of the form
\begin{align}
\label{eq_Glauberseries_generic}
   \sigma^{\rm SLL+G}
   \sim \frac{\alpha_s\,L}{\pi\,N_c}\,\sum_{\ell=1}^\infty\sum_{n=0}^\infty \,c_{\ell,n}\,w_\pi^\ell\,w^{n+\ell} \,,
\end{align}
with $c_{\ell,n}\propto (-1)^{n+\ell}$.
Phenomenologically, the net effects of the higher Glauber corrections turn out to be small for these processes.
Based on the corresponding results for the SLLs obtained in~\cite{Becher:2023mtx}, however, one would expect larger corrections from the Glauber series if gluons are present in the initial state.
As they transform in the adjoint representation, the color algebra is much more involved in this case and the operator basis derived in~\cite{Becher:2023mtx,Boer:2023jsy} must be extended.

To be precise, recall the factorization formula for the gap-between-jets cross section for a $2 \to M$ wide-angle jet process at hadron colliders~\cite{Balsiger:2018ezi,Becher:2021zkk,Becher:2023mtx},
\begin{align} \label{eq_cross_section_general}
	\sigma_{2\to M}(Q_0) = \int\!dx_1 \! \int\!dx_2 \! \sum_{m=2+M}^{\infty} \tr{\H_m(\{\underline{n}\},s,x_1,x_2,\mu) \otimes \W_m(\{\underline{n}\},Q_0,x_1,x_2,\mu)} \,.
\end{align}
Here, the hard functions $\H_m$ describe the underlying partonic $1+2 \to 3 + \dots + m$ scattering process and $\W_m$ are the low-energy matrix elements.
Both objects are operators in color-space~\cite{Catani:1996vz}.
The sum includes all partonic channels with given multiplicity $m$, the symbol $\otimes$ denotes an integration over the directions of the final-state partons and $\{\underline{n}\}\equiv\{n_1,\dots, n_m\}$.
For more details on the notation and the following discussion, we refer to~\cite{Becher:2021zkk,Becher:2023mtx,Boer:2023jsy}.
The generalization from fixed angular constraints to cross sections defined in terms of sequential jet clustering is also feasible~\cite{Becher:2023znt}.

To perform the resummation, one exploits the RG equations for the hard functions 
\begin{equation}
    \frac{d}{d\ln\mu}\,\H_m (\{\underline{n}\},s,\mu) = - \sum_{l=2+M}^m \H_l(\{\underline{n}\},s,\mu) \star \GammaH_{lm}(\{\underline{n}\},s,\mu) \,,
\end{equation}
whose formal solution can be expressed in terms of the path-ordered exponential 
\begin{equation} \label{eq_path_ordered_exponential}
    \bm{U}(\{\underline{n}\},s,\mu_h,\mu) = \mathbf{P} \exp\left[\int_{\mu}^{\mu_h}\frac{d\mu'}{\mu'} \GammaH(\{\underline{n}\},s,\mu')\right] .
\end{equation}
Note that the anomalous dimension $\GammaH$ is an operator in both color space and the infinite space of parton multiplicities.
This fact makes the evaluation of the solution a challenging task.
Its action on a hard function is defined through a series expansion which schematically reads 
\begin{align}
\label{eq_evolution_series}
    &\H(\mu_h) \star \bm{U}(\mu_h,\mu) \\*
    &= \H(\mu_h) + \int_\mu^{\mu_h}\!\frac{d\mu'}{\mu'}\,\H(\mu_h) \star \GammaH(\mu') 
     + \int_\mu^{\mu_h}\!\frac{d\mu'}{\mu'} \int_{\mu'}^{\mu_h}\!\frac{d\mu''}{\mu''}\,
     \H(\mu_h) \star \GammaH(\mu'') \star \GammaH(\mu') + \dots \,. \nonumber
\end{align}
For simplicity of the notation, the dependence of the hard functions and anomalous dimensions on the momentum fractions carried by the initial-state partons is omitted. 
The $\star$ symbol indicates a Mellin convolution over these momentum fractions.
The one-loop anomalous dimension $\GammaH$ can be found e.g.~in (2.13) of~\cite{Becher:2023mtx}.
For jet processes at lepton colliders, the anomalous dimension is known up to two-loop order~\cite{Becher:2021urs}.
The successive application of $\GammaH$ leads to the appearance of color traces with increasing complexity. 
When working to leading-logarithmic accuracy, however, two major simplifications can be exploited in the analysis.

First, as discussed in Section 3 of~\cite{Becher:2023mtx}, the one-loop result of the anomalous dimension $\GammaH$ can be expressed as $\GammaH=\GammaS+\GammaC$,\footnote{Note that the soft anomalous dimension $\GammaS$ has a trivial dependence on the momentum fractions and, therefore, the Mellin convolution evaluates to $1$.} where only the soft part 
\begin{equation} \label{eq_anomalous_dimension_soft_part}
    \GammaS=\frac{\alpha_s}{4\pi}\, \Big(\Gammac \ln \frac{\mu^2}{\mu_h^2} + \VG+\GammaBar \Big) + \mathcal{O}(\alpha_s^2)
\end{equation}
is relevant in the following.
In particular, for the Glauber series one considers arbitrarily many insertions of the soft-collinear emission operator $\Gammac$ and the Glauber operator $\VG$, but only a single insertion of the soft-emission operator $\GammaBar$.
The reason is that, as explained after~\eqref{eq_SLLgeneric}, the contributions of $\Gammac$ and $\VG$ are logarithmically enhanced.
Iterated insertions of these objects therefore determine the coefficients in the Glauber series~\eqref{eq_Glauberseries_generic}.
On the contrary, multiple insertions of $\GammaBar$ or $\GammaC$ lead to subleading logarithmic corrections.
Explicitly, the various objects in~\eqref{eq_anomalous_dimension_soft_part} are defined as
\begin{align} \label{eq_VGGammac}
    \Gammac &= \sum_{i=1,2} \, \gamma_0^{\rm cusp} \big[C_i\,\Id - \T_{i,L}\circ\T_{i,R}\,\delta(n_{k}-n_i) \big] \,,
    \nonumber\\
    \VG &= - 2i\pi\,\gamma_0^{\rm cusp}\,\big( \T_{1,L}\cdot\T_{2,L} - \T_{1,R}\cdot\T_{2,R} \big) \,, 
    \\
    \GammaBar &= 2 \sum_{(ij)} \left( \T_{i,L}\cdot\T_{j,L} + \T_{i,R}\cdot\T_{j,R} \right) \int\frac{d\Omega(n_k)}{4\pi}\,\overline{W}_{ij}^k - 4 \sum_{(ij)} \T_{i,L}\circ\T_{j,R}\,\overline{W}_{ij}^{k}\,\Theta_{\rm hard}(n_{k}) \,, \nonumber
\end{align}
where $\gamma_0^{\rm cusp} = 4$ is the one-loop coefficient of the cusp anomalous dimension and $C_i$ denotes the eigenvalue of the quadratic Casimir operator for the $i$-th parton.
The subtracted soft dipole $\overline{W}^{k}_{ij}$ (with collinear singularities removed) is defined as
\begin{align}
    \overline{W}_{ij}^k \equiv W_{ij}^k - \frac{1}{n_i\cdot n_k}\,\delta(n_i- n_k)  - \frac{1}{n_j\cdot n_k}\,\delta(n_j- n_k) \,; \qquad W^k_{ij} = \frac{n_i \cdot n_j}{n_i \cdot n_k \, n_j \cdot n_k} \,,
\end{align}
and $\Theta_{\rm hard}(n_k)$ restricts the direction $n_k$ of the emission to be inside the jet.
Note that $\VG$ is diagonal in multiplicity space while $\Gammac$ is an upper bidiagonal matrix.
Whereas these two operators only involve the color generators of the initial-state partons ($i=1,2$), the operator $\GammaBar$ contains color generators for all partons in the process.
In our formalism the hard functions $\H_m$ are density matrices, and the subscript $L\,(R)$ indicates that the color generators act on this matrix from the left (right), i.e.~on the amplitude (complex conjugate amplitude).
The structures $\Gammac$ and $\GammaBar$ also contain real-emission contributions, and the symbol $\circ$ indicates the emission of an additional gluon in the direction $n_k$, which technically implies an extension of the color space.
The $\delta$-function in $\Gammac$ then enforces this gluon to be collinear with respect to the light-like direction $n_i$ of its emitter.

Second, after evolving the hard functions down to the soft scale $\mu_s\sim Q_0$ using~\eqref{eq_path_ordered_exponential}, one can use the lowest-order matrix elements $\W_m(\mu_s)$ in the form
\begin{align}
\label{eq_LOlowenergyME}
    \W_m(\{\underline{n}\},Q_0,x_1,x_2,\mu_s) = f_1(x_1,\mu_s) \, f_2(x_2,\mu_s) \, \Id + \mathcal{O}(\alpha_s)\,,
\end{align}
with the standard parton distribution functions $f_i(x_i)$.

The above simplifications, in combination with the identities
\begin{align} \label{eq_prop_GammaC_GammaBar_VG}
    \com{\Gammac}{\GammaBar} = 0 \,, \qquad
    \tr{\H\,\Gammac\otimes\Id} = 0 \,, \qquad
    \tr{\H\,\VG\otimes\Id} = 0
\end{align}
imply that the coefficients of the Glauber series are associated with the the color traces~\cite{Boer:2023jsy}
\begin{equation} \label{eq_color_trace_general}
	C_{\{\underline{r}\}}^\ell \equiv \tr{\H_{2\to M} \left(\Gammac\right)^{r_1} \VG \left(\Gammac\right)^{r_2} \VG \dots \left(\Gammac\right)^{r_{2\ell-1}} \VG \left(\Gammac\right)^{r_{2\ell}} \VG \, \GammaBar \otimes \Id} \,,
\end{equation}
where $\H_{2\to M}$ are the Born-level hard functions.
An even number $2\ell$ of Glauber operators is required to yield a non-vanishing result, while a total number of $n\equiv\sum_{i=1}^{2\ell} r_i$ structures $\Gammac$ can be inserted as indicated above.
For $\ell=1$, this expression reduces to the color traces for the SLLs studied in~\cite{Becher:2021zkk,Becher:2023mtx}.
It is convenient to define the abbreviation
\begin{equation} \label{eq_H_abbreviation}
	\H \equiv \H_{2\to M} \left(\Gammac\right)^{r_1} \VG \left(\Gammac\right)^{r_2} \VG \dots \left(\Gammac\right)^{r_{2\ell-1}} \VG \left(\Gammac\right)^{r_{2\ell}} \,,
\end{equation}
in terms of which one can simplify the trace~\eqref{eq_color_trace_general} to~\cite{Becher:2021zkk,Becher:2023mtx}
\begin{equation} \label{eq_color_trace_first_Glauber}
	C_{\{\underline{r}\}}^\ell = 64i\pi \,\sum_{j>2} J_j \, if^{abc}\, \tr{\H \, \T_1^a\,\T_2^b\,\T_j^c \otimes \Id} \,.
\end{equation}
Here, $J_j$ denotes the integral over the soft dipoles with a veto $\Theta_{\text{veto}}(n_k)\equiv 1 - \Theta_{\rm hard}(n_k)$ enforcing the soft emission to the inside of the veto region,\footnote{Inside the veto region $W_{ij}^k$ and $\overline{W}_{ij}^k$ coincide since the emission cannot be collinear to the jets.}
\begin{equation} \label{eq_angular_integrals}
	J_j \equiv \int \frac{d\Omega(n_k)}{4\pi} \left(W_{1j}^k-W_{2j}^k\right) \Theta_{\text{veto}}(n_k)\,.
\end{equation}
The angular integration is implicit in the symbol $\otimes$ in~\eqref{eq_cross_section_general} for terms containing a real emission~\cite{Becher:2023mtx}.
In the following, $\otimes\,\Id$ is omitted inside all color traces for brevity.

The remainder of this paper is organized as follows:
The systematic reduction of the color traces~\eqref{eq_color_trace_general}, first for gluon-initiated, and then for quark-gluon-initiated processes, is described in Section~\ref{sec_colortrace}.
In Section~\ref{sec_resummation} the resummation of the reduced traces, combined with the nested scale integrals from~\eqref{eq_evolution_series} is performed.
Numerical estimates for the contribution of higher-order Glauber terms to all possible $2\to0$, $2\to1$ and $2\to2$ processes are presented in Section~\ref{sec_pheno}, before concluding in Section~\ref{sec_conclusions}.

%%%%%%%%%%%%%%%%%%%%%%%%%%%%%%%%%%%%%%%%%%%%%%%%%%%%%%%%%%%%%%%%%%
\section{Systematic reduction of the color traces}
\label{sec_colortrace}

The central task is the evaluation of the color traces~\eqref{eq_color_trace_general}.
For particles transforming in the fundamental representation, products of generators can always be reduced to a single generator or the identity matrix.
This feature was exploited in~\cite{Boer:2023jsy} to resum the Glauber series for quark-initated processes.
However, such a property does not exist for gluons, which transform in the adjoint representation.
Due to the plethora of $\Gammac$ and $\VG$ insertions, the color traces~\eqref{eq_color_trace_general} then involve color structures of seemingly arbitrary complexity. 
Remarkably, it is still possible to reduce the color trace in this case.
This is achieved by constructing a finite color basis that is closed under repeated applications of $\VG$ and $\Gammac$. 
This allows for a resummation of the Glauber series for initial states that feature gluons, thus extending the analysis in~\cite{Boer:2023jsy}.

\subsection{Construction of the color bases}
\label{subsec_bases}
Since the color traces~\eqref{eq_color_trace_general} contain only a single insertion of $\GammaBar$, and $\VG$ as well as $\Gammac$ only depend on the color generators $\T_{1,2}$ of the initial-state partons, at most one generator $\T_j$ of a final-state parton appears, see~\eqref{eq_color_trace_first_Glauber}. 
Hence, one can decompose all possible color structures into two distinct classes, corresponding to certain linear combinations of either
\begin{equation} \label{eq_general_types_color_structures}
	\zeta \, \bm{\mathcal{C}}_1 \, \widetilde{\bm{\mathcal{C}}}_2 \, \T_j \qquad \text{or} \qquad \zeta \, \bm{\mathcal{C}}_1 \, \widetilde{\bm{\mathcal{C}}}_2 \,,
\end{equation}
where the tilde indicates that these structures are not necessarily related by interchanging $1\leftrightarrow2$.
If the initial-state partons are either both (anti-)quarks or both gluons, these linear combinations are constrained by additional symmetries from relabeling particles $1\leftrightarrow2$. 
The objects $\bm{\mathcal{C}}_i$ and $\widetilde{\bm{\mathcal{C}}}_i$ are color-space matrices which contain products of color generators associated with parton $i$.
One the one hand, they carry two \emph{matrix} indices, i.e.~anti-fundamental or fundamental indices if parton $i$ is a quark or anti-quark, respectively, or adjoint indices if it is a gluon.
On the other hand, they also carry an open \emph{adjoint} index for each color generator.
Whereas the matrix indices are to be contracted with the hard function under the color trace, the open adjoint indices are contracted with $\zeta$, a color-space tensor of corresponding rank.
For example, the color structure in~\eqref{eq_color_trace_first_Glauber} is of the first class, with
\begin{equation}
	\zeta = if^{abc} \,, \qquad \bm{\mathcal{C}}_1 = \T_1^a \,, \qquad \widetilde{\bm{\mathcal{C}}}_2 = \T_2^b\,,
\end{equation}
and describes the soft emission of a gluon from final-state parton $j$.
In contrast, the second class in~\eqref{eq_general_types_color_structures}, without the additional generator $\T_j$, describes soft emissions originating from collinear gluons~\cite{Becher:2021zkk,Becher:2023mtx,Boer:2023jsy}.

One possible choice for the structures $\bm{\mathcal{C}}_1$ and $\widetilde{\bm{\mathcal{C}}}_2$ are symmetrized products of $SU(N_c)$ generators. Spelling out adjoint indices explicitly, they read
\begin{equation} \label{eq_color_basis_symmetrized_generators}
	\bm{\mathcal{C}}_i^{(k)a_1\dots a_k} = \frac{1}{k!} \sum_{\sigma\in S_k} \T_i^{a_{\sigma(1)}} \dots \T_i^{a_{\sigma(k)}} \,,
\end{equation}
where $\bm{\mathcal{C}}_i^{(0)}=\Id_i$ and the sum is over all permutations of $\{1,2,\dots,k\}$.
The open adjoint indices must be contracted with $\zeta$, which can be constructed from all combinations and permutations of the symbols $\delta^{a_1a_2}$, $if^{a_1a_2a_3}$ and $d^{a_1a_2a_3}$, for example
\begin{equation}
    \zeta^{(0)} = 1 \,, \qquad \zeta^{(2)a_1a_2} = \delta^{a_1a_2} \,, \qquad \zeta^{(3)a_1a_2a_3} \in\{ if^{a_1a_2a_3} \,, \, d^{a_1a_2a_3} \} \,,
\end{equation}
and so on.
Here, $d^{a_1a_2a_3}$ are the totally symmetric and traceless coefficients.
Note that higher-order $d$-symbols are recursively defined via $\delta^{a_1a_2}$ and $d^{a_1a_2a_3}$, see e.g.~\cite{Klein:1963}.

For the treatment of the SLLs, it suffices to consider symmetrized products of up to three generators, cf.~(6.36) and~(6.45) in~\cite{Becher:2023mtx}. 
However, inserting more and more Glauber operators in~\eqref{eq_color_trace_general} creates color structures that contain symmetrized products of more and more generators. 
Therefore, it is not possible to construct a \emph{finite} color basis valid for initial-state partons transforming under \emph{any} representation.
However, upon specifying a representation, one can always construct a finite basis, as will be shown for the fundamental and adjoint representations below. 
This is a somewhat unexpected result, given the complexity of the color algebra in the adjoint representation.

\subsubsection{Quark-initiated processes}
\label{subsec_basis_quark}
For the initial-state partons $i = 1,2$ being (anti-)quarks, one can use the relation
\begin{equation} \label{eq_product_two_generators_fundamental}
    \t_i^a \, \t_i^b = \frac{1}{2N_c} \, \delta^{ab} \, \Id_i + \frac{1}{2}\left(if^{abc} + \sigma_i \, d^{abc}\right) \t_i^c
\end{equation}
to construct a finite color basis for this case~\cite{Boer:2023jsy}.  
The color-space formalism implies that $(\t_i^a)_{\alpha_i\beta_i}=-(t^a)^T_{\alpha_i\beta_i}$, $\sigma_i=-1$ if the initial-state parton $i$ is a quark and $(\t_i^a)_{\alpha_i\beta_i}=(t^a)_{\alpha_i\beta_i}$, $\sigma_i=+1$ for an anti-quark, where $t^a$ are generators of the fundamental representation.
For the two classes of structures shown in~\eqref{eq_general_types_color_structures}, one now constructs all contractions of $\bm{\mathcal{C}}_1 \in \{ \Id_1, \t_1^a \}$ and $\widetilde{\bm{\mathcal{C}}}_2 \in \{\Id_2, \t_2^b\}$ with $\zeta \in \{\delta^{bc}, \delta^{ac}, if^{abc}, \sigma_1 d^{abc}, \sigma_2 d^{abc} \}$ for the first class and $\zeta \in \{1, \delta^{ab} \}$ for the second class of structures.
As a consequence of~\eqref{eq_product_two_generators_fundamental}, the $d$-symbols are always accompanied by $\sigma_{1,2}$. 

Note that the cross section is invariant under the relabeling $1\leftrightarrow2$. 
The structures from the first class always appear in combination with the angular integral $J_j$, where $j>2$, whereas the ones from the second class are accompanied by $J_{12} \equiv J_2$.
According to~\eqref{eq_angular_integrals}, these angular integrals transform as
\begin{align}
    J_j \to - J_j \,, \qquad J_{12} \to + J_{12}
\end{align}
under the exchange $1\leftrightarrow2$, it suffices to consider the anti-symmetric combinations
\begin{equation}
    if^{abc}\,\t_1^a\,\t_2^b\,\T_j^c \,, \qquad (\sigma_1-\sigma_2)\,d^{abc}\,\t_1^a\,\t_2^b\,\T_j^c \,, \qquad (\t_1 - \t_2)\cdot\T_j\,,
\end{equation}
in the former case, and the symmetric combinations
\begin{equation}
    \t_1\cdot\t_2 \,, \qquad \Id \,,
\end{equation}
in the latter case.
These five color structures form a basis for the case of quark-initiated processes, as has been explicitly verified in~\cite{Boer:2023jsy}.

\subsubsection{Gluon-initiated processes}
\label{subsec_color_basis_adjoint}
Consider next the situation where both initial-state partons are gluons, i.e.~transform in the adjoint representation.
In this case, the generators for parton $i=1,2$ can be expressed through the structure constants as
\begin{equation}
	(\T_i^a)_{a_ib_i} = -if^{aa_ib_i} \,.
\end{equation}
Correspondingly, the two matrix indices of the color structures $\bm{\mathcal{C}}_i$ and $\widetilde{\bm{\mathcal{C}}}_i$ in~\eqref{eq_general_types_color_structures}, that are not contracted with $\zeta$, are adjoint indices as well.

As already mentioned in Section~\ref{subsec_basis_quark}, the cross section~\eqref{eq_cross_section_general} is invariant under the relabeling $1\leftrightarrow2$, therefore it suffices to consider only anti-symmetric linear combinations
\begin{equation} \label{eq_Aj_structure}
	\Aj = \zeta \, (\bm{\mathcal{C}}_1 \, \widetilde{\bm{\mathcal{C}}}_2 - \bm{\mathcal{C}}_2 \, \widetilde{\bm{\mathcal{C}}}_1) \, \T_j
\end{equation}
and symmetric linear combinations
\begin{equation} \label{eq_S_structure}
	\S = \zeta \, (\bm{\mathcal{C}}_1 \, \widetilde{\bm{\mathcal{C}}}_2 + \bm{\mathcal{C}}_2 \, \widetilde{\bm{\mathcal{C}}}_1)
\end{equation}
as possible basis structures.
Combining $\zeta$ with the linear combinations of $\bm{\mathcal{C}}_i$ and $\widetilde{\bm{\mathcal{C}}}_i$ in~\eqref{eq_Aj_structure} yields a tensor with five open adjoint indices. 
Four of those correspond to the open \emph{matrix} indices, and the additional index is the one contracted with the stripped-off final-state generator $\T_j$.
Similarly, the tensor in~\eqref{eq_S_structure} carries only the four open matrix indices.

In order to determine the number of basis structures, one can express these tensors in terms of traces of generators in the fundamental representation.
Consider first the $\Aj$ structures and, therefore, tensors with five open indices.
Since the generators are traceless and normalized such that $\text{tr}(t^a t^b) = \frac12 \delta^{ab}$, the only allowed tensors consist of anti-symmetric (in $(a_1,b_1)\leftrightarrow(a_2,b_2)$) linear combinations of permutations of the traces
\begin{equation} \label{eq_fund_traces_5}
	\text{tr}(t^c \, t^{a_1} \, t^{b_1} \, t^{a_2} \, t^{b_2}) \quad\text{and}\quad \text{tr}(t^c \, t^{a_1} \, t^{b_1}) \, \delta^{a_2b_2}\,.
\end{equation}
To obtain a basis element, this linear combination is then multiplied by the left-over $\T_j^c$.
Due to the cyclicity of the trace, there are $4!+\binom{5}{3} \,2!=44$ permutations of~\eqref{eq_fund_traces_5}, $22$ of which are anti-symmetric.

One can apply the same arguments to the $\S$ structures, where now the only allowed tensors are symmetric linear combinations of permutations of
\begin{equation} \label{eq_fund_traces_4}
	\text{tr}(t^{a_1} \, t^{b_1} \, t^{a_2} \, t^{b_2})\quad\text{and}\quad\delta^{a_1b_1} \, \delta^{a_2b_2}\,.
\end{equation}
Whereas the $\frac{1}{2}\binom{4}{2}=3$ permutations of the second term are already symmetric, the $3!$ permutations of the first term allow for additional four symmetric linear combinations, yielding in total seven linearly independent $\S$ structures.

Since the linear combinations of~\eqref{eq_fund_traces_5} and~\eqref{eq_fund_traces_4} are not particularly suited to be used within the color-space formalism, it is advantageous to map them on objects that naturally appear in the adjoint representation.
To do so, one expresses the traces of fundamental generators in terms of $\delta^{ab}$, $if^{abc}$ and $d^{abc}$, which carry only adjoint indices, and rearranges the indices in such a way that one ends up with products of
\begin{equation} \label{eq_adjoint_matrices}
\begin{aligned}
	(\Id_i)_{a_ib_i} &\equiv \delta^{a_ib_i} \,, & (\F_i^a)_{a_ib_i} &\equiv -if^{aa_ib_i} \,, & (\DELTA_i^{\!ab})_{a_ib_i} &\equiv \delta^{aa_i}\delta^{bb_i} + \delta^{ba_i}\delta^{ab_i} \,,
	\\
	&& (\D_i^a)_{a_ib_i} &\equiv d^{aa_ib_i} \,, & (\NABLA_i^{ab})_{a_ib_i} &\equiv \delta^{aa_i}\delta^{bb_i} - \delta^{ba_i}\delta^{ab_i} \,,
\end{aligned}
\end{equation}
for initial-state partons $i=1,2$. These matrices fulfill the important relations~\cite{Haber:2019sgz}
\begin{equation}\label{eq_importantRelationsGluon}
	\begin{aligned}
		\com{\F_i^a}{\F_i^b} &= if^{abc} \F_i^c \,, &  \com{\F_i^a}{\D_i^b} &= \com{\D_i^a}{\F_i^b} = if^{abc} \D_i^c \,,
		\\
		\com{\D_i^a}{\D_i^b} &= if^{abc} \F_i^c - \frac{2}{N_c} \NABLA^{ab}_i \,,\qquad & \F_i^a \D_i^b + \F_i^b \D_i^a &= \D_i^a \F_i^b + \D_i^b \F_i^a = d^{abc} \F_i^c \,.
	\end{aligned}
\end{equation}
Using~\eqref{eq_importantRelationsGluon}, it is possible to reduce commutators of these matrices, mimicking the properties of ordinary generators of $SU(N_c)$.
The anti-commutators of these matrices, however, cannot be simplified in general, leading to a more complicated scenario than in the quark case.
Instead, the matrices satisfy the useful identity~\cite{Haber:2019sgz}
\begin{equation}\label{eq_UsefulIdentityGluonAnticom}
	\DELTA_i^{\!ab} + \frac{N_c}{2}\anticom{\F_i^a}{\F_i^b} + \frac{N_c}{2}\anticom{\D_i^a}{\D_i^b} = 2\delta^{ab}\, \Id_i + N_c\,d^{abc}\D_i^c \,.
\end{equation}
In the following, we choose to remove $\anticom{\D_i^a}{\D_i^b}$ using~\eqref{eq_UsefulIdentityGluonAnticom}. Remarkably, as shown below, it suffices to consider color structures constructed from the matrices~\eqref{eq_adjoint_matrices}, complemented by the two additional anti-commutators $\anticom{\F_i^a}{\F_i^b}$ and $\anticom{\F_i^a}{\D_i^b}$.
It is thus not necessary to include the symmetrized products as in~\eqref{eq_color_basis_symmetrized_generators} for $k\geq3$.

\begin{table}[t] 
	\centering
	\renewcommand{\arraystretch}{1.5}
	\centerline{
		\begin{tabular}{|l|l|l|l|} 
		      \hline
			\multicolumn{2}{|c|}{\cellcolor{Lightgray} Two contracted indices} & 
			\multicolumn{2}{c|}{\cellcolor{Lightgray} Four contracted indices}\\
			\hline
			$\Aj_{2,F}$ & $\left(\F_1 - \F_2\right) \cdot \T_j$
			& $\Aj_{4,F,\Delta}$ & $\left(\F_1^a \, \DELTA_2^{ab} - \mbox{\small $1\!\leftrightarrow\!2$}\right)\T_j^b$
			\\
			$\Aj_{2,D}$ & $\left(\D_1 - \D_2\right) \cdot \T_j$
			& $\Aj_{4,F,\nabla}$ & $\left(\F_1^a \, \NABLA_2^{ab} - \mbox{\small $1\!\leftrightarrow\!2$}\right)\T_j^b$
			\\ \hhline{--}
			\multicolumn{2}{c|}{\phantom{text}} & $\Aj_{4,F,FF}$ & $\left(\F_1^a \, \anticom{\F_2^a}{\F_2^b} - \mbox{\small $1\!\leftrightarrow\!2$}\right)\T_j^b$
			\\\hhline{--}
			\multicolumn{2}{|c|}{\cellcolor{Lightgray} Three contracted indices} & $\Aj_{4,F,FD}$ & $\left(\F_1^a \, \anticom{\F_2^a}{\D_2^b} - \mbox{\small $1\!\leftrightarrow\!2$}\right)\T_j^b$
			\\ \hhline{--}
			$\Aj_{3f,F,F}$ & $if^{abc} \, \F_1^a \, \F_2^b \, \T_j^c$
			& $\Aj_{4,D,\Delta}$ & $\left(\D_1^a \, \DELTA_2^{ab} - \mbox{\small $1\!\leftrightarrow\!2$}\right)\T_j^b$
			\\
			$\Aj_{3f,D,D}$ & $if^{abc} \, \D_1^a \, \D_2^b \, \T_j^c$
			& $\Aj_{4,D,\nabla}$ & $\left(\D_1^a \, \NABLA_2^{ab} - \mbox{\small $1\!\leftrightarrow\!2$}\right)\T_j^b$
			\\
			$\Aj_{3f,F,D}$  & $if^{abc} \left(\F_1^a \D_2^b - \F_2^a \D_1^b \right) \T_j^c$
			& $\Aj_{4,D,FF}$ & $\left(\D_1^a \, \anticom{\F_2^a}{\F_2^b} - \mbox{\small $1\!\leftrightarrow\!2$}\right)\T_j^b$
			\\
			$\Aj_{3d,F,D} $ & $d^{abc} \left(\F_1^a \D_2^b - \F_2^a \D_1^b \right) \T_j^c$
			& $\Aj_{4,D,FD}$ & $\left(\D_1^a \, \anticom{\F_2^a}{\D_2^b} - \mbox{\small $1\!\leftrightarrow\!2$}\right)\T_j^b$
			\\ \hline
			\multicolumn{4}{c}{}
			\\\hline
			\multicolumn{4}{|c|}{\cellcolor{Lightgray} Five contracted indices} \\ 
			\hline
            $\Aj_{5f,\Delta,\Delta}$ & $if^{abc} \, \DELTA_1^{ad} \, \DELTA_2^{bd} \, \T_j^c$ & &
			\\
			$\Aj_{5f,\nabla,\nabla}$ & $if^{abc} \, \NABLA_1^{ad} \, \NABLA_2^{bd} \, \T_j^c$ & &
			\\
			$\Aj_{5f,\Delta,\nabla}$ & $if^{abc} \left(\DELTA_1^{ad} \, \NABLA_2^{bd} - \mbox{\small $1\!\leftrightarrow\!2$}\right) \T_j^c$
			& $\Aj_{5d,\Delta,\nabla}$ & $d^{abc} \left(\DELTA_1^{ad} \, \NABLA_2^{bd} - \mbox{\small $1\!\leftrightarrow\!2$}\right) \T_j^c$
			\\
			$\Aj_{5f,\Delta,FF}$ & $if^{abc} \left(\DELTA_1^{ad} \, \anticom{\F_2^b}{\F_2^d} - \mbox{\small $1\!\leftrightarrow\!2$}\right) \T_j^c$
			& $\Aj_{5d,\Delta,FF}$ & $d^{abc} \left(\DELTA_1^{ad} \, \anticom{\F_2^b}{\F_2^d} - \mbox{\small $1\!\leftrightarrow\!2$}\right) \T_j^c$
			\\
			$\Aj_{5f,\Delta,FD}$ & $if^{abc} \left(\DELTA_1^{ad} \, \anticom{\F_2^b}{\D_2^d} - \mbox{\small $1\!\leftrightarrow\!2$}\right) \T_j^c$
			& $\Aj_{5d,\Delta,FD}$ & $d^{abc} \left(\DELTA_1^{ad} \, \anticom{\F_2^b}{\D_2^d} - \mbox{\small $1\!\leftrightarrow\!2$}\right) \T_j^c$
			\\
			$\Aj_{5f,\nabla,FF}$ & $if^{abc} \left(\NABLA_1^{ad} \, \anticom{\F_2^b}{\F_2^d} - \mbox{\small $1\!\leftrightarrow\!2$}\right) \T_j^c$
			& $\Aj_{5d,\nabla,FF}$ & $d^{abc} \left(\NABLA_1^{ad} \, \anticom{\F_2^b}{\F_2^d} - \mbox{\small $1\!\leftrightarrow\!2$}\right) \T_j^c$
			\\
			$\Aj_{5f,\nabla,FD}$ & $if^{abc} \left(\NABLA_1^{ad} \, \anticom{\F_2^b}{\D_2^d} - \mbox{\small $1\!\leftrightarrow\!2$}\right) \T_j^c$
			& $\Aj_{5d,\nabla,FD}$ & $d^{abc} \left(\NABLA_1^{ad} \, \anticom{\F_2^b}{\D_2^d} - \mbox{\small $1\!\leftrightarrow\!2$}\right) \T_j^c$
			\\ \hline	
	\end{tabular}
	}
	\caption{Possible anti-symmetric color structures featuring $\T_j$ for gluon-initiated processes.}
	\label{tab_Aj_structures}
\end{table}

\paragraph{$\Aj$ structures:}
The relations between traces of three and five fundamental generators, as they appear in~\eqref{eq_fund_traces_5}, and $\delta^{ab}$, $if^{abc}$ and $d^{abc}$ are
\begin{equation} \label{eq_fund_trace35}
\begin{split}
    \text{tr}(t^a\,t^b\,t^c) &= \frac{1}{4} (d^{abc} + if^{abc}) \,,
    \\
    \text{tr}(t^a\,t^b\,t^c\,t^d\,t^e) &= \frac{1}{8N_c} \, \delta^{cd} \, (d^{abe}+if^{abe}) + \frac{1}{8N_c} \, \delta^{ab} \, (d^{cde} + if^{cde})
    \\
    &\qquad +\frac{1}{16} (d^{abf}+if^{abf})(d^{cdg}+if^{cdg})(d^{efg}+if^{efg}) \,,
\end{split}
\end{equation}
which can easily be proven by applying~\eqref{eq_product_two_generators_fundamental}.
From the second relation it follows that one only needs to consider color structures $\Aj$ that contain at most three $f$- or $d$-symbols.
In total, there are $26$ such color structures, listed in Table~\ref{tab_Aj_structures}, where they are sorted by the number of contracted indices, i.e.~by the indices of $\zeta$ in~\eqref{eq_Aj_structure}.
All other contractions can either be reduced to those listed in the table or vanish by symmetry.
For the possible structures of $\Aj_3$-type \emph{not} listed in Table~\ref{tab_Aj_structures}, one can apply
\begin{equation} \label{eq_simplifications_A3}
\begin{aligned}
	if^{abc} \, \NABLA_i^{ab} &= -2 \F_i^c \,, \qquad
	&
	d^{abc} \, \anticom{\F_i^a}{\F_i^b} &= N_c \,\D_i^c \,,
	\\
	d^{abc} \, \DELTA_i^{ab} &= 2 \D_i^c \,,
	&
	d^{abc} \, \anticom{\F_i^a}{\D_i^b} &= \left(\frac{N_c^2-4}{N_c}\right) \F_i^c \,,
\end{aligned}
\end{equation}
to relate these operators to the $\Aj_2$. 
In principle, it is also possible to construct $\Aj_4$-type structures by contracting with $\zeta\in\{f^{abe}f^{cde},d^{abe}d^{cde},if^{abe}d^{cde}\}$.
However, contractions with two additional anti-commutators cannot contribute, as they already contain four $f$- or $d$-symbols and one can directly simplify
\begin{equation}
\begin{aligned}
    f^{ace}f^{bde} \DELTA_i^{ab} &= \anticom{\F_i^c}{\F_i^d} \,, \qquad
    &
    f^{ace}f^{bde} \NABLA_i^{ab} &= if^{cde} \F_i^e \,,
	\\
	d^{ace}d^{bde} \DELTA_i^{ab} &= \anticom{\D_i^c}{\D_i^d} \,, \qquad
	&
	d^{ace}d^{bde} \NABLA_i^{ab} &= if^{cde} \F_i^e - \frac{2}{N_c} \NABLA_i^{cd} \,,
	\\
	if^{ace}d^{bde} \DELTA_i^{ab} &= if^{cde} \D_i^e \,, \qquad
	&
	if^{ace}d^{bde} \NABLA_i^{ab} &= \anticom{\F_i^c}{\D_i^d} \,.
\end{aligned}
\end{equation}
Recall that the anti-commutator $\anticom{\D_i^c}{\D_i^d}$ can be eliminated by means of~\eqref{eq_UsefulIdentityGluonAnticom}.

Table~\ref{tab_Aj_structures} contains $26$ color structures, but the basis consists only of $22$ structures, following the argument given around~\eqref{eq_fund_traces_5}.
Indeed, there are four non-trivial relations
\begin{equation} \label{eq_Aj_structures_linear_dependence}
\begin{split}
	\Aj_{5f,\nabla,FD} &= - \Aj_{4,F,FD} \,,
	\\
	\Aj_{5d,\Delta,FF} &= \Aj_{4,D,FF} \,,
	\\
	\Aj_{5f,\Delta,FD} &= 2\, \Aj_{3f,F,D} + \Aj_{5d,\nabla,FF} \,,
	\\
	\Aj_{5d,\Delta,FD} &=  - \frac{8}{N_c}\, \Aj_{2,F} - 4\, \Aj_{3d,F,D} + \Aj_{4,F,FF} - \Aj_{4,D,FD} - \Aj_{5f,\nabla,FF} \,,
\end{split}
\end{equation}
which can be used to reduce the full set to $22$ basis structures.
With the explicit basis at hand, one can construct the isomorphism to the basis~\eqref{eq_fund_traces_5}, for example
\begin{equation}
\begin{split}
    \text{tr}(t^c \, t^{a_1} \, t^{b_1} \, t^{a_2} \, t^{b_2}) - (1\!\leftrightarrow\!2) \quad &\Leftrightarrow \quad \frac{1}{8} \big(\Aj_{3f,F,F} + \Aj_{3f,D,D} - \Aj_{3f,F,D}\big) \,,
    \\
    \text{tr}(t^c \, t^{a_1} \, t^{b_1}) \, \delta^{a_2b_2} - (1\!\leftrightarrow\!2) \quad &\Leftrightarrow \quad \frac{1}{4} \big(\Aj_{2,D} - \Aj_{2,F}\big) \,,
\end{split}
\end{equation}
which allows one to express the traces of fundamental generators in terms of the basis structures listed in Table~\ref{tab_Aj_structures}.

\begin{table}[t] 
	\centering
	\renewcommand{\arraystretch}{1.5}
    \centerline{
		\begin{tabular}{|l|l|l|l|}
			\hline
			\multicolumn{2}{|c|}{\cellcolor{Lightgray} No contracted indices} & 
			\multicolumn{2}{c|}{\cellcolor{Lightgray} Two contracted indices}\\
			\hline
			$\S_{0}$ & $\Id$
			& $\S_{2,F,F}$ & $\F_1 \cdot \F_2$
			\\ \cline{1-2}
			\multicolumn{2}{c|}{}& $\S_{2,D,D}$ & $\D_1 \cdot \D_2$
			\\
			\multicolumn{2}{c|}{}& $\S_{2,F,D}$ & $\F_1 \cdot \D_2 + \F_2 \cdot \D_1$
			\\ \cline{3-4}
			\multicolumn{4}{c}{}
			\\ \hline
			\multicolumn{4}{|c|}{\cellcolor{Lightgray} Four contracted indices} \\ 
			\hline
			$\S_{4,\Delta,\Delta}$ & $\DELTA_1^{ab} \, \DELTA_2^{ab}$
			& $\S_{4,\Delta,FF}$ & $\DELTA_1^{ab} \, \anticom{\F_2^a}{\F_2^b} + \DELTA_2^{ab} \, \anticom{\F_1^a}{\F_1^b}$
			\\
			$\S_{4,\nabla,\nabla}$ & $\NABLA_1^{ab} \, \NABLA_2^{ab}$
			& $\S_{4,\Delta,FD}$ & $\DELTA_1^{ab} \, \anticom{\F_2^a}{\D_2^b} + \DELTA_2^{ab} \, \anticom{\F_1^a}{\D_1^b}$
			\\
			& & \textcolor{Gray}{$\S_{4,\nabla,FD}$} & \textcolor{Gray}{$\NABLA_1^{ab} \, \anticom{\F_2^a}{\D_2^b} + \NABLA_2^{ab} \, \anticom{\F_1^a}{\D_1^b}$}
			\\ \hline	
	\end{tabular}
	}
	\caption{Possible symmetric color structures without $\T_j$ for gluon-initiated processes. Even though the individual terms in $\S_{4,\nabla,FD}$ are non-zero, they vanish in the symmetric combination (gray).}
	\label{tab_S_structures}
\end{table}

\paragraph{$\S$ structures:} The trace of four fundamental generators, appearing in~\eqref{eq_fund_traces_4}, fulfills
\begin{equation} \label{eq_fund_trace4}
    \text{tr}(t^a\,t^b\,t^c\,t^d) = \frac{1}{4N_c} \,\delta^{ab}\delta^{cd} + \frac{1}{8} (d^{abe}d^{cde} - f^{abe}f^{cde} + if^{abe}d^{cde} + d^{abe} if^{cde}) \,,
\end{equation}
which can also be proven by applying~\eqref{eq_product_two_generators_fundamental} twice.
This immediately implies that one only needs to consider $\S$ structures with up to two $f$- or $d$-symbols.
There are eight such color structures, listed in Table~\ref{tab_S_structures}, sorted by the number of contracted indices, i.e.~by the indices of $\zeta$ in~\eqref{eq_S_structure}.
All other contractions can either be reduced to these or vanish by symmetry, i.e.~all possible $\S_3$-type structures can directly be reduced to $\S_2$ using~\eqref{eq_simplifications_A3}. 
Since the basis of $\S$ structures contains only seven elements, as argued around~\eqref{eq_fund_traces_4}, one non-trivial relation among the structures in Table~\ref{tab_S_structures} exist, which reads
\begin{equation} \label{eq_S_structures_linear_dependence}
	\S_{4,\Delta,FD} = 2\, \S_{2,F,D} \,.
\end{equation}
Hence, one reduces the full set to seven basis structures.
Again, it is possible to find the isomorphism to the basis~\eqref{eq_fund_traces_4}.
For example,
\begin{equation}
\begin{split}
	\text{tr}(t^{a_1} \, t^{b_1} \, t^{a_2} \, t^{b_2}) \quad &\Leftrightarrow \quad \frac{1}{4N_c} \, \S_0 + \frac{1}{8} \big(\S_{2,F,F} + \S_{2,D,D} - \S_{2,F,D}\big) \,,
	\\
	\delta^{a_1b_1}\,\delta^{a_2b_2} \quad &\Leftrightarrow \quad \frac{1}{4} \big(\S_{4,\Delta,\Delta} + \S_{4,\nabla,\nabla}\big) \,,
\end{split}
\end{equation}
where both terms on the left-hand side are already symmetric under $1\leftrightarrow2$.

Remarkably, this construction yields a finite basis containing $22+7$ color structures for the infinite Glauber series of gluon-initiated processes. 

\subsubsection{Quark-gluon-initiated processes}
\label{subsec_color_basis_mixed}
As the third scenario, consider an initial state consisting of one (anti-)quark and one gluon.
Without loss of generality, parton $1$ is always assumed to be the (anti-)quark, i.e.
\begin{equation}
	\T_1^a = \t_1^a \qquad \text{and} \qquad \T_2^a = \F_2^a \,.
\end{equation}
It is now no longer possible to restrict the form of color structures by symmetry arguments under the relabeling $1\leftrightarrow2$.
Hence, one can only distinguish two types of structures
\begin{equation} \label{eq_Oj_O_structure}
\begin{split}
    \Oj &= \zeta \, \bm{\mathcal{C}}_1 \, \widetilde{\bm{\mathcal{C}}}_2 \, \T_j \,,
    \\
    \O &= \zeta \, \bm{\mathcal{C}}_1 \, \widetilde{\bm{\mathcal{C}}}_2 \,.
\end{split}
\end{equation}
Relation~\eqref{eq_product_two_generators_fundamental} implies that $\bm{\mathcal{C}}_1$ can always be reduced to $\Id_1$ or $\t_1^a$.
It is again possible to determine the number of basis structures by constructing a basis out of fundamental traces, as explained around~\eqref{eq_fund_traces_5} and~\eqref{eq_fund_traces_4} for the case of gluon-initiated processes.
Starting with structures $\Oj$, the number of adjoint indices of $\zeta\,\widetilde{\bm{\mathcal{C}}}_2$ is three if combined with $\Id_1$ and four if accompanied by $\t_1^a$.
The only allowed tensors thus consist of all permutations of
\begin{equation} \label{eq_fund_traces_Oj}
    \text{tr}(t^c\,t^{a_2}\,t^{b_2}) \quad\text{and}\quad \text{tr}(t^c\,t^a\,t^{a_2}\,t^{b_2}) \,,\, \delta^{ca} \, \delta^{a_2b_2} \,,
\end{equation}
multiplied by $\Id_1$ and $\t_1^a$, respectively.
To obtain a basis element, these traces are then combined with the left-over $\T_j^c$.
Similarly, for the structures $\O$, the number of adjoint indices of $\zeta\,\widetilde{\bm{\mathcal{C}}}_2$ is two or three, and the allowed tensors are permutations of
\begin{equation} \label{eq_fund_traces_O}
    \delta^{a_2b_2} \quad\text{and}\quad \text{tr}(t^a\,t^{a_2}\,t^{b_2}) \,,
\end{equation}
combined with $\Id_1$ and $\t_1^a$, respectively.
Taking the cyclicity of the trace into account, one finds $2!+3!+\frac{1}{2}\binom{4}{2}=11$ basis structures $\Oj$ and $1+2!=3$ structures $\O$.
\begin{table}[t] 
	\centering
	\renewcommand{\arraystretch}{1.5}
	\begin{tabular}{|l|l|l|l|l|l|}
		\hline
		\multicolumn{2}{|c|}{\cellcolor{Lightgray} Two contracted indices} & 
		\multicolumn{2}{c|}{\cellcolor{Lightgray} Three contracted indices} &
		\multicolumn{2}{c|}{\cellcolor{Lightgray} Four contracted indices} \\
		\hline
		$\Oj_{2,F}$ & $\F_2\cdot\T_j$ & $\Oj_{3f,F}$ & $if^{abc}\,\t_1^a\,\F_2^b\,\T_j^c$ & $\Oj_{4,\Delta}$ & $\t_1^a\,\DELTA_2^{ab}\,\T_j^b$ \\
		$\Oj_{2,D}$ & $\sigma_1\,\D_2\cdot\T_j$ & $\Oj_{3f,D}$ & $\sigma_1\,if^{abc}\,\t_1^a\,\D_2^b\,\T_j^c$ & $\Oj_{4,\nabla}$ & $\t_1^a\,\NABLA_2^{ab}\,\T_j^b$ \\
		$\Oj_{2,t}$ & $\t_1\cdot\T_j$ & $\Oj_{3d,F}$ & $\sigma_1\,d^{abc}\,\t_1^a\,\F_2^b\,\T_j^c$ & $\Oj_{4,FF}$ & $\t_1^a\,\anticom{\F_2^a}{\F_2^b}\,\T_j^b$ \\ \cline{1-2}
		\multicolumn{2}{c|}{} & $\Oj_{3d,D}$ & $d^{abc}\,\t_1^a\,\D_2^b\,\T_j^c$ & $\Oj_{4,FD}$ & $\sigma_1\,\t_1^a\,\anticom{\F_2^a}{\D_2^b}\,\T_j^b$
		\\ \cline{3-6}
	\end{tabular}
	\\[0.5cm]
	\begin{tabular}{|l|l|l|l|}
	    \hline
	    \multicolumn{2}{|c|}{\cellcolor{Lightgray} No contracted indices} & 
		\multicolumn{2}{c|}{\cellcolor{Lightgray} Two contracted indices} \\ \hline
		$\O_0$ & $\Id$ & $\O_{2,F}$ & $\t_1\cdot\F_2$ \\ \cline{1-2}
		\multicolumn{2}{c|}{} & $\O_{2,D}$ & $\sigma_1\,\t_1\cdot\D_2$
		\\ \cline{3-4}
    \end{tabular}
	\caption{Possible color structures with and without $\T_j$ for quark-gluon-initiated processes.}
	\label{tab_Oj_O_structures}
\end{table}

Again, the structures~\eqref{eq_fund_traces_Oj} and~\eqref{eq_fund_traces_O} are not suited to be used within the color-space formalism and are therefore mapped onto objects that naturally appear in this context.
An obvious choice for $\bm{\mathcal{C}}_1$ is to be either $\Id_1$ or $\t_1^a$.
For the color structure $\widetilde{\bm{\mathcal{C}}}_2$ of parton~$2$, we choose the adjoint matrices~\eqref{eq_adjoint_matrices} as well as the anti-commutators $\anticom{\F_2^a}{\F_2^b}$, $\anticom{\F_2^a}{\D_2^b}$.
Combining them to $\Oj$ and $\O$ color structures results in $11+3$ structures, listed in Table~\ref{tab_Oj_O_structures}.
For each $d$-symbol one includes a factor $\sigma_1$ (recall that $\sigma_1^2=1$), thus allowing for a simultaneous treatment of parton 1 being a quark or an anti-quark.
By the same arguments as given in Section~\ref{subsec_color_basis_adjoint}, no additional color structures are necessary.
Therefore, these $14$ color structures are linearly independent and constitute a basis for the color algebra of (anti-)quark-gluon-initiated processes relevant for the Glauber series.

\subsection{Reduction to the bases}
\label{subsec_reduction}
With the color bases at hand, the reduction of the color traces~\eqref{eq_color_trace_general} is now straightforward.
One first computes the action of $\Gammac$ and $\VG$ on all basis structures and then expresses the reduced color traces as linear combinations of the basis structures with coefficients depending on $\ell$ and $\{\underline{r}\}$.

\subsubsection{Gluon-initiated processes}
\label{subsec_reduction_adjoint}
The Glauber operator acts on a color structure $\Aj$ as
\begin{equation} \label{eq_action_VG_Aj}
	\tr{\H \, \VG \Aj_i} = -8i\pi \tr{\H \, \com{\Aj_i}{\F_1\cdot\F_2}} \,.
\end{equation}
The action of $\VG$ on $\S$ structures can be calculated in a similar way. It turns out that all commutators vanish, and hence
\begin{equation} \label{eq_action_VG_S_zero}
	\S_i \overset{\VG}{\longrightarrow} 0 \,.
\end{equation}

\begin{figure}[t]
	\centering
	\includegraphics[scale=1]{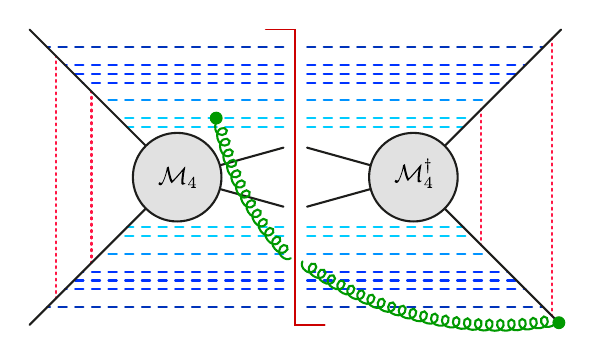}
	\caption{Graphical illustration of a contribution to the color traces $C_{\{\underline{r}\}}^{\ell}$ involving the $\S$ structures, relevant for $M=2$ jet production. The soft wide-angle emission from the operator $\GammaBar$ is shown in green and can be attached as explained in the main text. Glauber terms $\VG$ are shown with red dotted lines and collinear emissions $\Gammac$ as dashed blue lines, lighter blue colors indicate ``earlier” collinear emissions (closer to the hard function). This particular diagram shows a contribution to $C_{\{4,2,6,2\}}^2$ which involves $r_1=4$ emissions before the first, $r_2=2$ before the second, $r_3=6$ before the third and $r_4=2$ emissions before the last Glauber phase.}
	\label{fig_S_strutures}
\end{figure}

The action of $\Gammac$ differs for basis structures with and without $\T_j$. While the virtual part $\Vc_{1,2}\sim\Id$ acts trivially on both types, the real part $\Rc_{1,2}$ can map structures $\Aj$ to $\S$, but not vice versa.
The action on an $\S$ structure is
\begin{equation} \label{eq_action_Gammac_S}
	\tr{\H \, \Gammac \, \S_i} = 4 \Big[ (C_1+C_2) \, \tr{\H \, \S_i} - \tr{\H\,\F_1^a\,\S_i\,\F_1^a} - \tr{\H\,\F_2^a\,\S_i\,\F_2^a}\Big] ,
\end{equation}
where $C_1=C_2=C_A=N_c$ for gluon-initiated processes.
The action on an $\Aj$ structure is more complicated.
One finds
\begin{align} \label{eq_action_Gammac_Aj}
	\sum_{j>2} J_j \, \tr{\H \, \Gammac \, \Aj_i} = 4 \sum_{j>2}{}\strut^\prime J_j &\Big[(C_1+C_2) \tr{\H \, \Aj_i} - \tr{\H\,\F_1^a\,\Aj_i\,\F_1^a} - \tr{\H\,\F_2^a\,\Aj_i\,\F_2^a} \Big]
	\nonumber\\
	{}+ 4\,J_{12} &\Big[\tr{\H\,\F_1^a\,\Aj_i\,\F_1^b} - \tr{\H\,\F_2^a\,\Aj_i\,\F_2^b} \Big] \Big|_{\T_j^c\to-if^{cab}} \,,
\end{align}
where the prime on the sum indicates that the gluon whose emission is described by $\Rc_{1,2}$ is excluded.
The term in the second line of~\eqref{eq_action_Gammac_Aj} describes the emission of a wide-angle soft gluon off a collinear gluon emitted from parton $1$ or $2$, see Figure~\ref{fig_S_strutures}.
More details on the relevant color algebra can be found in~\cite{Becher:2023mtx}.

\begin{table}[t]
	\centering
	\renewcommand{\arraystretch}{1.5}
	\begin{tabular}{|lllll||ll|}
	    \hline
		\cellcolor{Lightgray}$\VG\,\GammaBar$ & \cellcolor{Lightgray}$(\VG)^2\,\GammaBar$ & \cellcolor{Lightgray}$(\VG)^3\,\GammaBar$ & \cellcolor{Lightgray}$(\VG)^4\,\GammaBar$ & \cellcolor{Lightgray}$(\VG)^5\,\GammaBar$ & \multicolumn{2}{c|}{\cellcolor{Lightgray}never appear} \\ \hline
		$\Aj_{3f,F,F}$ & $\Aj_{2,F}$ & \textcolor{Gray}{$\Aj_{3f,F,F}$} & \textcolor{Gray}{$\Aj_{2,F}$} & \textcolor{Gray}{$\Aj_{3f,F,F}$} & $\Aj_{3f,F,D}$ & $\Aj_{2,D}$ \\
		& $\Aj_{3d,F,D}$ & $\Aj_{3f,D,D}$ & \textcolor{Gray}{$\Aj_{3d,F,D}$} & \textcolor{Gray}{$\Aj_{3f,D,D}$} & $\Aj_{4,D,\nabla}$ & $\Aj_{4,D,\Delta}$ \\
		& $\Aj_{4,F,\Delta}$ & $\Aj_{4,F,\nabla}$ & \textcolor{Gray}{$\Aj_{4,F,\Delta}$} & \textcolor{Gray}{$\Aj_{4,F,\nabla}$} & $\Aj_{5d,\Delta,\nabla}$ & $\Aj_{4,D,FF}$ \\
		& $\Aj_{4,F,FF}$ & $\Aj_{5f,\Delta,\Delta}$ & \textcolor{Gray}{$\Aj_{4,F,FF}$} & \textcolor{Gray}{$\Aj_{5f,\Delta,\Delta}$} & $\Aj_{5d,\nabla,FF}$ & $\Aj_{4,F,FD}$ \\
		& & $\Aj_{5f,\Delta,FF}$ & $\Aj_{4,D,FD}$ & \textcolor{Gray}{$\Aj_{5f,\Delta,FF}$} & & \\
		& & & $\Aj_{5f,\Delta,\nabla}$ & $\Aj_{5f,\nabla,\nabla}$ & & \\
		& & & $\Aj_{5f,\nabla,FF}$ & $\Aj_{5d,\nabla,FD}$ & & \\ \hline
		& $\S_0$ & & \textcolor{Gray}{$\S_0$} & & & $\S_{2,F,D}$ \\
		& $\S_{2,F,F}$ & & \textcolor{Gray}{$\S_{2,F,F}$} & & & \\
		& $\S_{2,D,D}$ & & \textcolor{Gray}{$\S_{2,D,D}$} & & & \\
		& $\S_{4,\Delta,\Delta}$ & & \textcolor{Gray}{$\S_{4,\Delta,\Delta}$} & & & \\
		& $\S_{4,\Delta,FF}$ & & \textcolor{Gray}{$\S_{4,\Delta,FF}$} & & & \\
		& & & $\S_{4,\nabla,\nabla}$ & & & \\ \hline\hline
		$1+0$ & $4+5$ & $5+0$ & $7+6$ & $7+0$ & $4+0$ & $4+1$ \\ \hline
	\end{tabular}
	\caption{Color basis for gluon-initiated processes. The different columns list the basis structures appearing after pulling out a given number of $\VG$ operators from $\H$ in~\eqref{eq_color_trace_first_Glauber}. Structures appearing for the first time are indicated in black, whereas structures that have appeared already before are indicated in gray. For a given number of Glauber operators, application of $\Gammac$ does not create new structures containing $\T_j$ but does create the $\S$ structures shown in the lower half. The one exception from this rule is $\Aj_{4,F,FF}$, which only appears after applying $\Gammac$ to $(\VG)^2\,\GammaBar$. The two last columns list $8+1$ basis structures that never appear starting out from~\eqref{eq_color_trace_first_Glauber}.}
	\label{tab_basis_adjoint}
\end{table}

Table~\ref{tab_basis_adjoint} summarizes how the basis structures constructed in Section~\ref{subsec_color_basis_adjoint} are related under the mapping of Glauber operators and collinear anomalous dimensions.
It is convenient to order the structures that appear by successively pulling out factors of $\VG$ and $\Gammac$ from $\H$ in~\eqref{eq_color_trace_first_Glauber}, i.e.~by application onto $\Aj_{3f,F,F}$, in the following way:
\begin{equation} \label{eq_def_X_vector_adjoint}
	\bm{X} \equiv \Big(\underbrace{\sum_{j>2}J_j \,\Aj_{3f,F,F},\dots}_{\text{7 structures}} \, , \underbrace{\sum_{j>2}J_j \,\Aj_{2,F},\dots}_{\text{7 structures}} \, , \underbrace{J_{12}\,\S_0,\dots \vphantom{\sum_{j>2}}}_{\text{6 structures}} \Big)^T \,.
\end{equation}
Here the first seven structures emerge from an odd number of $\VG$ insertions and are the ones from the fifth column of Table~\ref{tab_basis_adjoint}.
The remaining thirteen structures emerge from an even number of insertions and are given in the fourth column.\footnote{Note that the operators from the first three columns are included in the fourth and fifth one as well.}
Applying further insertions of $\VG$ does not create new color structures.
Besides these 20 basis structures there are nine structures which are not generated by the operators $\VG$ and $\Gammac$ in the color trace (shown in the right portion of the table).
In this \emph{physical}\footnote{Physical in the sense that it only contains the $14+6$ color structures appearing in the Glauber series calculation and not the $8+1$ structures from the last two columns in Table~\ref{tab_basis_adjoint}.} basis one can write for an arbitrary hard function $\H$
\begin{equation} \label{eq_Gammac_VG_matrices}
\begin{split}
    \tr{\H\,\VG\,\bm{X}_i} &= \sum_{\tilde{\imath}} \left(V^G\right)_{i\tilde{\imath}} \tr{\H\,\bm{X}_{\tilde{\imath}}} \,,
    \\
    \tr{\H\,\Gammac\,\bm{X}_i} &= \sum_{\tilde{\imath}} \left(\Gamma^c\right)_{i\tilde{\imath}} \tr{\H\,\bm{X}_{\tilde{\imath}}} \,,
\end{split}
\end{equation}
with $20\times20$ matrices $V^G$ and $\Gamma^c$. 
Here the implicit sum over $j$ contained in the definition of $\bm{X}$ in~\eqref{eq_def_X_vector_adjoint} is different for the left-hand side and the right-hand side of the second equation, as explained below~\eqref{eq_action_Gammac_Aj}. 
The matrix $V^G$ can be written in block form as
\begin{equation} \label{eq_VG_decomposition_adjoint} 
	V^G = (i\pi)\,(4N_c)
	\begin{pmatrix}
		\phantom{l}0_{7\times7}\phantom{l} & \nu^{(j)} & 0_{7\times6} \\[0.3cm]
		\tilde{\nu}^{(j)} & \phantom{l}0_{7\times7}\phantom{l} & 0_{7\times6} \\[0.3cm]
		0_{6\times7} & 0_{6\times7} & \phantom{l}0_{6\times6}\phantom{l}
	\end{pmatrix} .
\end{equation}
The simple form of this matrix is the motivation for the ordering of the color structures as presented in~\eqref{eq_def_X_vector_adjoint}.
Here the two $0_{7\times7}$ matrices on the diagonal reflect the fact that the $\Aj$ structures can be split into two distinct subsets, each appearing only for an even or odd number of Glauber-operator insertions.
The last six zero rows indicate that $\S$ structures are mapped onto zero, cf.~\eqref{eq_action_VG_S_zero}, whereas the two $0_{7\times6}$ matrices in the last column indicate that $\VG$ does not create $\S$ structures when acting on $\Aj$.
In an analogous way, the $r$-th power of the collinear anomalous dimension matrix can be decomposed as
\begin{equation} \label{eq_Gammac_decomposition_adjoint}
	\left(\Gamma^c\right)^r = (4N_c)^r
	\begin{pmatrix}
		\phantom{l}\tilde{\gamma}^{(j)}(r)\phantom{l} & 0_{7\times7} & 0_{7\times6} \\[0.3cm]
		0_{7\times7} & \phantom{l}\gamma^{(j)}(r)\phantom{l} & \lambda(r) \\[0.3cm]
		0_{6\times7} & 0_{6\times7} & \phantom{l}\gamma(r)\phantom{l}
	\end{pmatrix} .
\end{equation}
The positions of the non-zero entries indicate that $\Aj$ structures appearing for an odd number of $\VG$ insertions do not mix with those appearing for an even number or with $\S$ structures. 
Likewise, $\Gammac$ can only map $\Aj$ structures appearing for an even number of $\VG$ insertions onto $\S$ structures, cf.~\eqref{eq_action_Gammac_S} and~\eqref{eq_action_Gammac_Aj}.
By consistency, $\lambda(0)=0_{7\times6}$ and $\tilde{\gamma}^{(j)}(0)=\gamma^{(j)}(0)=1_{7\times7}$ as well as $\gamma(0)=1_{6\times6}$.
The submatrices of $V^G$ and $\Gamma^c$ have been calculated using \texttt{ColorMath}~\cite{Sjodahl:2012nk}and are given in Appendix~\ref{app_Matrix_representation}.

It is now possible to rewrite the color trace~\eqref{eq_color_trace_first_Glauber} as
\begin{equation}
\begin{split}
	C_{\{\underline{r}\}}^{\ell} &= 64i\pi \,\sum_{i} \varsigma_i^{(1|0)} \, \tr{\H \, \bm{X}_i} \,,
	\\
	&= 64i\pi \,\sum_{i,\tilde{\imath}} \varsigma_i^{(1|0)} \left[\left(\Gamma^c\right)^{r_{2\ell}} V^G \left(\Gamma^c\right)^{r_{2\ell-1}} \dots V^G \left(\Gamma^c\right)^{r_1}\right]_{i\tilde{\imath}}  \tr{\H_{2\to M}\,\bm{X}_{\tilde{\imath}}} \,,
	\\
	&\equiv 64i\pi \,\sum_{\tilde{\imath}} \varsigma_{\tilde{\imath}}^{(2\ell|r_1,\dots,r_{2\ell})} \, \tr{\H_{2\to M}\,\bm{X}_{\tilde{\imath}}} \,,
\end{split}
\end{equation}
where $\varsigma^{(1|0)}=(1,0,\dots,0)$.
The $\H$ in the first line is defined in~\eqref{eq_H_abbreviation}.
Since only an even number of Glauber-operator insertions is physically relevant, it is possible to rewrite the last line as
\begin{equation} \label{eq_color_trace_adjoint}
\begin{split}
	C_{\{\underline{r}\}}^{\ell} = \frac{16}{N_c}\,(-\pi^2)^{\ell} \,(4N_c)^{n+2\ell} \bigg\{\sum_{j=3}^{2+M} J_j &\sum_{i\in I^{(j)}} c_i^{(2\ell|r_1,\dots,r_{2\ell})} \, \tr{\H_{2\to M}\,\Aj_i} 
	\\
	 + J_{12} &\hspace{.4em}\sum_{i\in I} d_i^{(2\ell|r_1,\dots,r_{2\ell})} \, \tr{\H_{2\to M}\,\S_i}\bigg\} ,
\end{split}
\end{equation}
where $I^{(j)}$ and $I$ contain the $7+6$ basis structures from the fourth column of Table~\ref{tab_basis_adjoint}. Here $n=\sum_{i=1}^{2\ell} r_i$ is the total number of $\Gammac$ insertions.
The coefficients in~\eqref{eq_color_trace_adjoint} are defined by
\begin{equation}
	\varsigma^{(2\ell|r_1,\dots,r_{2\ell})} \equiv (4N_c)^{n+2\ell-1} \, (i\pi)^{2\ell-1} \Big(\tilde{c}^{(2\ell|r_1,\dots,r_{2\ell})} \,,\, c^{(2\ell|r_1,\dots,r_{2\ell})} \,,\, d^{(2\ell|r_1,\dots,r_{2\ell})} \Big) ,
\end{equation}
in analogy with~\eqref{eq_def_X_vector_adjoint}. This relation also holds for an odd number of Glauber-operator insertions, i.e.~$\ell\in\mathbb{N}/2$, and by consistency one has $\tilde{c}^{(1|0)}=(1,0,0,0,0,0,0)$, whereas $c^{(1|0)}$ and $d^{(1|0)}$ are zero vectors. Using the submatrices from the decompositions~\eqref{eq_VG_decomposition_adjoint} and~\eqref{eq_Gammac_decomposition_adjoint}, the relevant coefficients in~\eqref{eq_color_trace_adjoint} are found to be
\begin{equation} \label{eq_coefficients_adjoint}
\begin{split}
	c^{(2\ell|r_1,\dots,r_{2\ell})} &= \tilde{c}^{(1|0)} \, \tilde{\gamma}^{(j)}(r_{2\ell}) \, \nu^{(j)} \bigg(\prod_{i=2}^{\ell} \gamma^{(j)}(r_{2i-1}) \, \tilde{\nu}^{(j)} \, \tilde{\gamma}^{(j)}(r_{2i-2}) \, \nu^{(j)}\bigg)\, \gamma^{(j)}(r_1) \,,
	\\
	d^{(2\ell|r_1,\dots,r_{2\ell})} &= \tilde{c}^{(1|0)} \, \tilde{\gamma}^{(j)}(r_{2\ell}) \, \nu^{(j)} \bigg(\prod_{i=2}^{\ell} \gamma^{(j)}(r_{2i-1}) \, \tilde{\nu}^{(j)} \, \tilde{\gamma}^{(j)}(r_{2i-2}) \, \nu^{(j)}\bigg)\, \lambda(r_1) \,.
\end{split}
\end{equation}
The left-most product is trivial, $\tilde{c}^{(1|0)} \, \tilde{\gamma}^{(j)}(r_{2\ell}) = \tilde{c}^{(1|0)}$, and therefore
the coefficients are independent of $r_{2\ell}$ and the color trace~\eqref{eq_color_trace_adjoint} depends on $r_{2\ell}$ only through $n$. 
Note that the coefficients are independent of $\gamma(r)$, as $\VG$ maps all $\S$ structures onto zero, cf.~\eqref{eq_action_VG_S_zero}.
Comparing~\eqref{eq_coefficients_adjoint} to the corresponding result for quark-initiated processes, one notices that the coefficients in the gluon case are determined by a matrix product instead of a product of scalar functions, cf.~the discussion around (3.14) in~\cite{Boer:2023jsy}.

\subsubsection{Quark-gluon-initiated processes}
\label{subsec_reduction_mixed}
\begin{table}[t]
	\centering
	\renewcommand{\arraystretch}{1.5}
	\begin{tabular}{|llll|}
	    \hline
		\cellcolor{Lightgray}$\VG\,\GammaBar$ & \cellcolor{Lightgray}$(\VG)^2\,\GammaBar$ & \cellcolor{Lightgray}$(\VG)^3\,\GammaBar$ & \cellcolor{Lightgray}$(\VG)^4\,\GammaBar$ \\ \hline
		$\Oj_{3f,F}$ & $\Oj_{2,F}$ & \textcolor{Gray}{$\Oj_{3f,F}$} & \textcolor{Gray}{$\Oj_{2,F}$} \\
		& $\Oj_{2,t}$ & $\Oj_{3f,D}$ & \textcolor{Gray}{$\Oj_{2,t}$} \\
		& $\Oj_{3d,F}$ & $\Oj_{4,\nabla}$ & \textcolor{Gray}{$\Oj_{3d,F}$} \\
		& $\Oj_{3d,D}$ & & \textcolor{Gray}{$\Oj_{3d,D}$} \\
		& $\Oj_{4,\Delta}$ & & \textcolor{Gray}{$\Oj_{4,\Delta}$} \\
		& $\Oj_{4,FF}$ & & \textcolor{Gray}{$\Oj_{4,FF}$} \\
		& & & $\Oj_{2,D}$ \\
		& & & $\Oj_{4,FD}$ \\ \hline
		& $\O_0$ & & \textcolor{Gray}{$\O_0$} \\
		& $\O_{2,F}$ & & \textcolor{Gray}{$\O_{2,F}$} \\
		& & & $\O_{2,D}$ \\ \hline\hline
		$1+0$ & $6+2$ & $3+0$ & $8+3$ \\ \hline
	\end{tabular}
	\caption{Color basis for quark-gluon-initiated processes. The different columns list the basis structures appearing after pulling out a given number of $\VG$ operators from $\H$ in~\eqref{eq_color_trace_first_Glauber}. Structures appearing for the first time are indicated in black, whereas structures appeared already before are indicated in gray. For a given number of Glauber phases, $\Gammac$ does not create new structures containing $\T_j$ but creates the $\O$ structures shown in the lower half.}
	\label{tab_basis_mixed}
\end{table}
The action of the Glauber operator on color structures $\Oj$ can be calculated as in~\eqref{eq_action_VG_Aj}, one just has to replace $\F_1\to\t_1$.
Similar to~\eqref{eq_action_VG_S_zero} one finds for quark-gluon-initiated processes
\begin{equation} \label{eq_action_VG_O_zero}
	\O_i \overset{\VG}{\longrightarrow} 0 \,.
\end{equation}
The action of $\Gammac$ on color structures for quark-gluon-initiated processes can be calculated in analogy to~\eqref{eq_action_Gammac_S} and~\eqref{eq_action_Gammac_Aj}, but with $C_1=C_F$ and $C_2=C_A=N_c$.

Table~\ref{tab_basis_mixed} summarizes how the basis structures constructed in Section~\ref{subsec_color_basis_mixed} are related under the mapping of $\VG$ and $\Gammac$.
Again, it is convenient to order the basis structures as
\begin{equation} \label{eq_def_X_vector_mixed}
	\bm{X} \equiv \Big(\underbrace{\sum_{j>2}J_j \,\Oj_{3f,F},\dots}_{\text{3 structures}} \,, \underbrace{\sum_{j>2}J_j \,\Oj_{2,F},\dots}_{\text{8 structures}} \,, \underbrace{J_{12}\,\O_0,\dots\vphantom{\sum_{j>2}}}_{\text{3 structures}} \Big)^T ,
\end{equation}
where the first three structures emerge from an odd number of $\VG$ insertions and are the ones from the third column of Table~\ref{tab_basis_mixed}.
The remaining eleven structures emerge from an even number of insertions and are given in the fourth column.\footnote{Note that the operators from the first two columns are included in the last two as well.} 
Applying more Glauber operators does not give rise to new structures. Decomposing $V^G$ and $\Gamma^c$ as in~\eqref{eq_Gammac_VG_matrices}, one finds
\begin{equation} \label{eq_VG_decomposition_mixed}
	V^G = (i\pi)\,(4N_c)
	\begin{pmatrix}
		\phantom{l}0_{3\times3}\phantom{l} & \nu^{(j)} & 0_{3\times3} \\[0.3cm]
		\tilde{\nu}^{(j)} & \phantom{l}0_{8\times8}\phantom{l} & 0_{8\times3} \\[0.3cm]
		0_{3\times3} & 0_{3\times8} & \phantom{l}0_{3\times3}\phantom{l}
	\end{pmatrix}
\end{equation}
and
\begin{equation} \label{eq_Gammac_decomposition_mixed}
	\left(\Gamma^c\right)^r = (4N_c)^r
	\begin{pmatrix}
		\phantom{l}\tilde{\gamma}^{(j)}(r)\phantom{l} & 0_{3\times8} & 0_{3\times3} \\[0.3cm]
		0_{8\times3} & \phantom{l}\gamma^{(j)}(r)\phantom{l} & \lambda(r) \\[0.3cm]
		0_{3\times3} & 0_{3\times8} & \phantom{l}\gamma(r)\phantom{l}
	\end{pmatrix} .
\end{equation}
Here the positions of the non-zero submatrices reflect the same general properties as in~\eqref{eq_VG_decomposition_adjoint} and~\eqref{eq_Gammac_decomposition_adjoint} and they are given in Appendix~\ref{app_Matrix_representation}.

For quark-gluon-initiated processes the color traces~\eqref{eq_color_trace_general} can also be reduced to a form analogous to~\eqref{eq_color_trace_adjoint}, in which $I^{(j)}$ and $I$ now contain the $8+3$ basis structures from the last column of Table~\ref{tab_basis_mixed} and one has to replace $\Aj\to\Oj$ as well as $\S\to\O$.
The coefficients again fulfill~\eqref{eq_coefficients_adjoint} with $\tilde{c}^{(1|0)}=(1,0,0)$ whereas $c^{(1|0)}$ and $d^{(1|0)}$ are still zero vectors.

%%%%%%%%%%%%%%%%%%%%%%%%%%%%%%%%%%%%%%%%%%%%%%%%%%%%%%%%%%%%%%%%%%
\section{Resummation}
\label{sec_resummation}

The contribution of the Glauber series, consisting of the SLLs plus all higher-order Glauber terms, to a partonic cross section for a $2\to M$ jet process can be written as~\cite{Boer:2023jsy}
\begin{equation} \label{eq_rlsums}
   \hat{\sigma}_{2\to M}^{\rm SLL+G}(Q_0) 
   = \sum_{\ell=1}^{\infty} \sum_{\{\underline{r}\}} I_{\{\underline{r}\}}^{\ell}(\mu_h,\mu_s)\,
    C_{\{\underline{r}\}}^{\ell}
   \equiv \sum_{\ell=1}^{\infty} \sum_{r_1=0}^{\infty} \sum_{r_2=0}^{\infty} \ldots \sum_{r_{2\ell}=0}^{\infty} 
    \hat{\sigma}_{\{\underline{r}\}}^{\ell} \,.
\end{equation}
In addition to the color traces $C_{\{\underline{r}\}}^{\ell}$, one needs to determine the nested scale integrals $I_{\{\underline{r}\}}^{\ell}$ arising from the expansion of the path-ordered exponential in~\eqref{eq_evolution_series} to obtain the coefficients of the Glauber series.
These integrals are process-independent and, therefore, one can use the results from~\cite{Boer:2023jsy}.
In the strict double-logarithmic approximation (with the counting $\pi = |\ln(-1)| \sim L$) the running of the strong coupling $\alpha_s(\mu)$ can be neglected and one has
\begin{equation}
\label{eq_nested_integrals_evaluated}
    I_{\{\underline{r}\}}^{\ell}(\mu_h,\mu_s) 
    = \left(\frac{\alpha_s(\bar{\mu})}{4\pi}\right)^{2\ell+n+1} 
     \frac{(-2)^n\,L^{2n+2\ell+1}}{(2n+2\ell)(2n+2\ell+1)}\,\prod_{k=1}^{2\ell}\,
     \frac{(2\sum_{i=1}^{k-1}r_i+k-3)!!}{(2\sum_{i=1}^{k}r_i+k-1)!!} \,,
\end{equation}
where now $L=\ln(\mu_h/\mu_s)$, and $(-2)!!\equiv (-1)!!\equiv 1$.
Here $\bar{\mu}$ is an arbitrary fixed reference scale between $\mu_s$ and $\mu_h$.

Combining the coefficients of the color traces given in~\eqref{eq_color_trace_adjoint} with the nested scale integrals~\eqref{eq_nested_integrals_evaluated} results in the master formula for gluon-initiated processes
\begin{equation} \label{eq_master_formula}
\begin{split}
    \hat{\sigma}_{\{\underline{r}\}}^{\ell} 
    &= \frac{4\alpha_s(\bar{\mu}) L}{\pi N_c}  \frac{2^{n} \left(-w\right)^{\ell+n} w_\pi^{\ell} }{(2n+2\ell)(2n+2\ell+1)} \, \prod_{k=1}^{2\ell} \frac{(2\sum_{i=1}^{k-1}r_i+k-3)!!}{(2\sum_{i=1}^{k}r_i+k-1)!!} \\
    &\quad \times \bigg\{\sum_{j=3}^{2+M} J_j \sum_{i\in I^{(j)}} c_i^{(2\ell|r_1,\dots,r_{2\ell})} \, \tr{\H_{2\to M}\,\Aj_i} + J_{12} \sum_{i\in I} d_i^{(2\ell|r_1,\dots,r_{2\ell})} \, \tr{\H_{2\to M}\,\S_i}\bigg\} \,,
\end{split}
\end{equation}
with the $\mathcal{O}(1)$ expansion parameters $w$ and $w_\pi$ as defined below~\eqref{eq_SLLgeneric}.
The quark-gluon-initiated processes follow from this result using the replacements $\Aj\to\Oj$ and $\S\to\O$, see Section~\ref{subsec_reduction_mixed}.

The structure of this master formula is fairly similar to the quark case, cf.~(4.7) of~\cite{Boer:2023jsy}.
In particular, the alternating-sign behavior of the sum over $\ell$ has the effect of reducing the numerical impact of the SLL contributions to the partonic cross sections.
Additionally, the product of the fraction of double factorials quickly decreases for $\ell \to \infty$.
The expression in terms of multiple infinite sums can -- to the best of our knowledge -- not be evaluated in closed analytic form.
In~\cite{Boer:2023jsy} approximate analytic expressions were derived by restricting the number of insertions of the collinear anomalous dimension $\Gammac$ to not more than two.

An alternative way of organizing the Glauber series avoids the nested multi-sums by treating the Glauber operator as a perturbation, keeping the full exponential involving $\Gammac$ (or vice versa). 
Due to the identities~\eqref{eq_prop_GammaC_GammaBar_VG}, the form of the relevant color traces in~\eqref{eq_color_trace_general} requires the structure $\VG \, \GammaBar$ next to the $\otimes$ symbol, which implies that the path-ordered exponential~\eqref{eq_path_ordered_exponential} can be expressed as
\begin{equation} \label{eq_cross_section_Glauber_path_exp}
    \hat{\sigma}_{2\to M}^{\rm SLL+G} = \int_{\mu_s}^{\mu_h} \frac{d\mu}{\mu} \frac{\alpha_s(\mu)}{4\pi} \int_{\mu_s}^{\mu} \frac{d\mu'}{\mu'} \frac{\alpha_s(\mu')}{4\pi} \, \tr{\H_{2\to M}(\mu_h) \, \bm{U}^{\rm SLL+G}(\mu_h,\mu) \, \VG \, \GammaBar \otimes \Id} \,.
\end{equation}
Here the relevant evolution operator
\begin{equation} \label{eq_USLLG}
    \bm{U}^{\rm SLL+G}(\mu_h,\mu) \equiv \mathbf{P} \exp\left[\int_{\mu}^{\mu_h}\frac{d\mu'}{\mu'} \, \frac{\alpha_s(\mu')}{4\pi}\Big(\Gammac \, \ln\frac{\mu^{\prime\,2}}{\mu_h^2} + \VG\Big)\right]
\end{equation}
is restricted to the structures that are resummed to all orders.
Since the commutator $\com{\Gammac}{\VG}$ does not vanish this path-ordered exponential is a complicated object that cannot be expressed through simple functions of $\Gammac$ and $\VG$.
However, it is straightforward to expand~\eqref{eq_USLLG} in the Glauber operator, keeping the full dependence on $\Gammac$. This leads to an expansion of the form
\begin{equation} \label{eq_path_exp_expansion_VG}
    \hat{\sigma}_{2\to M}^{\rm SLL+G} = \hat{\sigma}_{2\to M}^{\rm SLL} + \hat{\sigma}_{2\to M}^{\rm SLL+G}\big|_{\ell=2} + \hat{\sigma}_{2\to M}^{\rm SLL+G}\big|_{\ell=3} + \dots \,,
\end{equation}
where the individual terms containing a total of $2\ell_0$ Glauber insertions are given by
\begin{equation}
\begin{split}
    \hat{\sigma}_{2\to M}^{\rm SLL+G}\big|_{\ell=\ell_0} &= \int_{\mu_s}^{\mu_h} \! \mathcal{D}^k\!\mu \: \langle \H_{2\to M}(\mu_h) \, \bm{U}^c(\mu_h,\mu_1) \, \VG \, \bm{U}^c(\mu_1,\mu_2) \, \VG \dots \\
    &\hspace{2.2cm} \times \bm{U}^c(\mu_{k-2},\mu_{k-1}) \, \VG \, \GammaBar \otimes \Id \rangle \,.
\end{split}
\end{equation}
Here, $k=2\ell_0+1$, the nested scale integral is defined by
\begin{equation}
    \int_{\mu_s}^{\mu_h} \! \mathcal{D}^k\!\mu \equiv \int_{\mu_s}^{\mu_h} \!\frac{d\mu_1}{\mu_1} \, \frac{\alpha_s(\mu_1)}{4\pi} \int_{\mu_s}^{\mu_1} \!\frac{d\mu_2}{\mu_2} \, \frac{\alpha_s(\mu_2)}{4\pi} \, \dots \int_{\mu_s}^{\mu_{k-1}} \!\frac{d\mu_k}{\mu_k} \, \frac{\alpha_s(\mu_k)}{4\pi} \,,
\end{equation}
and $\bm{U}^c$ is the matrix-valued Sudakov evolution operator, given by
\begin{equation}
    \bm{U}^c(\mu_2,\mu_1) \equiv \exp\bigg[\int_{\mu_1}^{\mu_2} \frac{d\mu}{\mu} \, \frac{\alpha_s(\mu)}{4\pi} \, \Gammac \ln\frac{\mu^2}{\mu_h^2}\bigg] \simeq \exp\bigg[\frac{\alpha_s(\bar{\mu})}{4\pi} \, \Gammac \left( \ln^2\frac{\mu_2}{\mu_h} - \ln^2\frac{\mu_1}{\mu_h} \right) \bigg] \,,
\end{equation}
where in the last step the integral is evaluated for a fixed coupling $\alpha_s(\bar{\mu})$. 
The first term in~\eqref{eq_path_exp_expansion_VG} represents the SLL series
\begin{equation} \label{eq_SLL_resummed_cross_section}
    \hat{\sigma}_{2\to M}^{\rm SLL} = \int_{\mu_s}^{\mu_h} \! \mathcal{D}^3\!\mu \: \tr{\H_{2\to M}(\mu_h) \, \bm{U}^c(\mu_h,\mu_1) \, \VG \, \bm{U}^c(\mu_1,\mu_2) \, \VG \, \GammaBar \otimes \Id} \,.
\end{equation}
Solving these scale integrals is highly non-trivial.
In the approximation where the running of the strong coupling is ignored the SLL contribution~\eqref{eq_SLL_resummed_cross_section} evaluates to a linear combination of Kamp\'e de F\'eriet functions~\cite{Becher:2023mtx}.
For higher Glauber terms the expressions become even more cumbersome and obtaining closed analytic results is beyond the scope of this article. 

Alternatively, one can expand~\eqref{eq_cross_section_Glauber_path_exp} in the number of $\Gammac$ insertions
\begin{equation} \label{eq_path_exp_expansion_Gammac}
    \hat{\sigma}_{2\to M}^{\rm SLL+G} = \hat{\sigma}_{2\to M}^{\rm SLL+G}\big|_{n=0} + \hat{\sigma}_{2\to M}^{\rm SLL+G}\big|_{n=1} + \hat{\sigma}_{2\to M}^{\rm SLL+G}\big|_{n=2} + \dots \,,
\end{equation}
i.e.~including all Glauber phases but only a finite number of collinear emissions.
Similar to above, the individual terms with $n_0$ insertions of $\Gammac$ are
\begin{align} \label{eq_path_exp_expansion_Gammac_terms}
    \hat{\sigma}_{2\to M}^{\rm SLL+G}\big|_{n=n_0} &= \int_{\mu_s}^{\mu_h} \! \mathcal{D}^k\!\mu \: \langle \H_{2\to M}(\mu_h) \, \bm{U}^G(\mu_h,\mu_1) \Big( \Gammac \ln\frac{\mu_1^2}{\mu_h^2} \Big) \bm{U}^G(\mu_1,\mu_2) \Big( \Gammac \ln\frac{\mu_2^2}{\mu_h^2} \Big) \dots \nonumber\\
    &\hspace{2.2cm} \times \bm{U}^G(\mu_{k-2},\mu_{k-1}) \, \VG \, \GammaBar \otimes \Id \rangle \,,
\end{align}
where now $k=n_0+2$ and
\begin{equation}
    \bm{U}^G(\mu_2,\mu_1) \equiv \exp\bigg[\int_{\mu_1}^{\mu_2} \frac{d\mu}{\mu} \, \frac{\alpha_s(\mu)}{4\pi} \, \VG \bigg] \simeq \exp\bigg[ \frac{\alpha_s(\bar{\mu})}{4\pi} \VG  \ln \frac{\mu_2}{\mu_1} \bigg] \,.
\end{equation}
A similar strategy to include higher-order Glauber operators through exponentiation in parton showers was investigated in~\cite{Nagy:2019rwb}.

In the approximation where the running of the coupling is ignored, one obtains for gluon-initiated processes
\begin{equation} \label{eq_resummedVG_0}
	\hat{\sigma}_{2\to M}^{\rm SLL+G}\big|_{n=0} 
	= \frac{4\alpha_s(\bar{\mu}) L}{\pi N_c} \, \sum_{k=1}^4 \left(\frac{\sin{v_{\pi,k}\sqrt{ww_{\pi}}}}{v_{\pi,k}\sqrt{ww_{\pi}}} - 1\right) \sum_{j=3}^{2+M} J_j \sum_{i\in I^{(j)}} \big(c_{\pi,k}^{(0)}\big)_i \, \tr{\H_{2\to M}\,\Aj_i} \,.
\end{equation}
Here, the eigenvalues of $\VG$ are
\begin{align}
	v_{\pi,1} &= 1 \,, & v_{\pi,2} &= \frac{2}{N_c} \,, & v_{\pi,3} &= \frac{N_c+2}{N_c} \,, & v_{\pi,4} &= \frac{N_c-2}{N_c} \,.
\end{align}
The coefficient vectors $c_{\pi,k}^{(0)}$ are given by
\begin{equation} \label{eq_VG_expansion_coeff}
	c_{\pi,k}^{(0)} = \frac{1}{v_{\pi,k}^{2}} \left(\tilde{c}^{(1|0)} \, \mathcal{V}_k \, \nu^{(j)}\right) ,
\end{equation}
where the matrix $\nu^{(j)}$ and the vector $\tilde{c}^{(1|0)}$ can be found in Section~\ref{subsec_reduction_adjoint}, and the matrices $\mathcal{V}_k$ are given in Appendix~\ref{app_resummation}.
A completely analogous result also holds for quark-gluon-initiated processes.
In this case one needs to replace $\Aj\to\Oj$ and sum only over the three eigenvalues
\begin{align}
    v_{\pi,2} &= \frac{2}{N_c} \,, & v_{\pi,3} &= \frac{N_c+1}{N_c} \,, & v_{\pi,4} &= \frac{N_c-1}{N_c} \,.
\end{align}
The coefficient vector is still obtained from~\eqref{eq_VG_expansion_coeff} with the corresponding objects for quark-gluon initiated processes given in Section~\ref{subsec_reduction_mixed} and in Appendix~\ref{app_resummation}.
The analogue of~\eqref{eq_resummedVG_0} for (anti-)quark-initiated processes was already presented in~\cite{Boer:2023jsy}. 
The second term in the $\Gammac$ expansion~\eqref{eq_path_exp_expansion_Gammac} is more complicated and can also be found in Appendix~\ref{app_resummation}.

%%%%%%%%%%%%%%%%%%%%%%%%%%%%%%%%%%%%%%%%%%%%%%%%%%%%%%%%%%%%%%%%%%
\section{Phenomenological implications}
\label{sec_pheno}

To estimate the size of higher-order Glauber exchanges, one can analyze their contribution to the total partonic cross section relative to the Born cross section
\begin{equation} \label{eq_normalization_hard_function}
    \hat{\sigma}_{2\to M} = \tr{\H_{2\to M} \otimes \Id}
\end{equation}
for different $2\to2$, $2\to1$, and $2\to0$ processes.
For $2\to 2$ processes, we restrict ourselves to small-angle scattering, in which case only a single color structure contributes to the amplitude.
The leading-order hard functions can then be expressed as
\begin{equation}
\label{eq_H_tensor}
    \H_{2\to M} = \tr{\H_{2\to M}} \; \mathcal{T}_{\alpha_1\dots\alpha_{2+M}} \, \mathcal{T}^{\dagger}_{\beta_1\dots\beta_{2+M}} \,,
\end{equation}
where $\mathcal{T}_{\alpha_1 \dots \alpha_{2+M}}$ denotes the color tensor associated with the amplitude $|\mathcal{M}_{2+M}\rangle$, and $\mathcal{T}^\dagger_{\beta_1\dots\beta_{2+M}}$ the tensor associated with the complex conjugate amplitude $\langle\mathcal{M}_{2+M}|$.
The indices $\alpha_i$ and $\beta_i$ are fundamental indices of $SU(N_c)$ if parton $i$ is a (anti-)quark, or adjoint indices if it is a gluon.
Note that the trace on the right-hand side of~\eqref{eq_H_tensor} does not contain the angular integrations from the~$\otimes$ symbol.

All numerical results presented below use two-loop running of the strong coupling with $\alpha_s(M_Z)=0.118$.
The ``all-order'' result of the Glauber series contains all terms in the double sum up to $\ell=3$ (corresponding to six $\VG$ insertions) and $n=20$ insertions of $\Gammac$, which gives sufficient numerical accuracy for the cases considered below.
The effect of the two-loop cusp anomalous dimension is included through the replacement~\cite{Becher:2023mtx}
\begin{equation}
    \alpha_s(\bar{\mu}) \to \left(1+\frac{\gamma_1^{\text{cusp}}}{\gamma_0^{\text{cusp}}}\,\frac{\alpha_s(\bar{\mu})}{4\pi}\right)\alpha_s(\bar{\mu}) \,.
\end{equation}

\subsection[Numerical estimates for \texorpdfstring{$2\to 2$}{2->2} processes]{Numerical estimates for $\bm{2\to 2}$ processes}
\label{subsec_2to2}
\begin{figure}[t]
    \centering
    \includegraphics[scale=1]{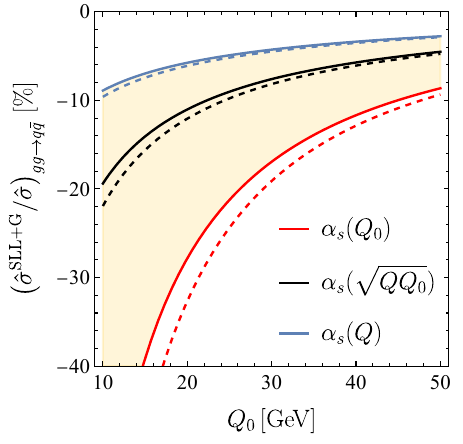}
    \includegraphics[scale=1]{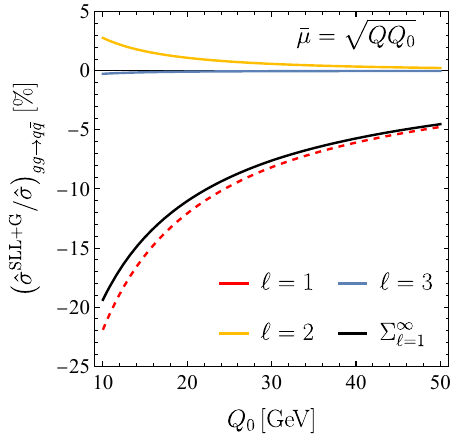}
    \\
    \includegraphics[scale=1]{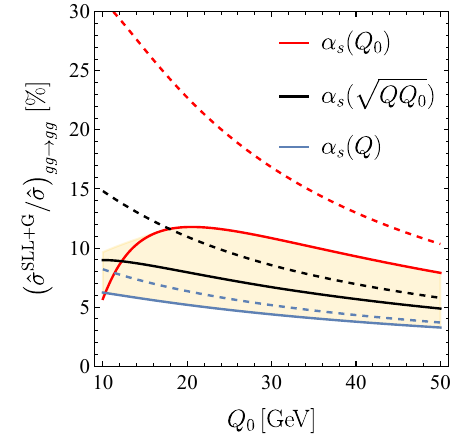}
    \includegraphics[scale=1]{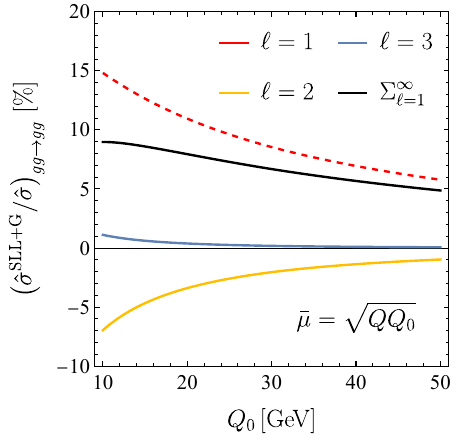}
    \caption{Numerical estimates of the impact of the Glauber series on $gg\to q\bar{q}$ (top row) and $gg\to gg$ (bottom row) small-angle scattering as a function of the jet-veto scale $Q_0$. The figures on the left show the all-order result obtained with three different choices for the scale $\bar{\mu}$ in the coupling. The figures on the right show the contributions of individual Glauber-phase pairs for $\bar{\mu}=\sqrt{QQ_0}$, with all large double logarithms resummed. Dashed lines show the corresponding curves in the SLL approximation ($\ell=1$). In all plots $Q=1$\,TeV and $\Delta Y=2$.}
    \label{fig_gg_2}
\end{figure}
For $2\to2$ processes, the leading-order hard functions have, in general, a non-trivial dependence on the kinematic variables. 
In the small-angle limit, these processes are dominated by the $t$- or $u$-channel exchange.
For $gg\to q\bar{q}$ and $gg\to gg$ scattering the cross section is symmetric, and only the $t$-channel diagrams are considered in the following.
In that case, the hard functions are  
\begin{equation} \label{eq_Hgg}
\begin{split}
    \H_{gg\to q\bar{q}} &= \tr{\H_{gg\to q\bar{q}}} \, \frac{1}{C_F^2 N_c} \, (t^{a_1}t^{a_2})_{\alpha_3\alpha_4} (t^{b_2}t^{b_1})_{\beta_4\beta_3} \,,
    \\
    \H_{gg\to gg} &= \tr{\H_{gg\to gg}} \, \frac{1}{N_c^2(N_c^2-1)} \, f^{a_1a_3a}f^{a_2a_4a}f^{b_1b_3b}f^{b_2b_4b} \,.
\end{split}
\end{equation}
However, the $qg \to qg$ cross section differs between the forward-scattering ($t$-channel dominated) and backward-scattering ($u$-channel dominated) limits. 
In the following, only forward scattering is considered, and the respective hard function reads
\begin{equation}
    \H_{qg\to qg} = \tr{\H_{qg\to qg}} \, \frac{2}{N_c(N_c^2-1)} \, f^{a_2a_4a} (t^a)_{\alpha_3\alpha_1} \, f^{b_2b_4b} (t^b)_{\beta_1\beta_3} \,.
\end{equation}
Evaluating the color traces of these hard functions with the basis elements $\Aj$, $\S$ and $\Oj$, $\O$, respectively, we can determine the different terms in the Glauber series~\eqref{eq_master_formula}.
For a central rapidity gap of width $\Delta Y$, the angular integrals~\eqref{eq_angular_integrals} evaluate to $J_{12}=\Delta Y$ as well as $J_3=-\Delta Y$ and $J_4=+\Delta Y$ for forward scattering~\cite{Becher:2023mtx}.
In gluon-initiated processes only the difference $(J_4-J_3)$ appears.

\begin{figure}[t]
    \centering
    \includegraphics[scale=1]{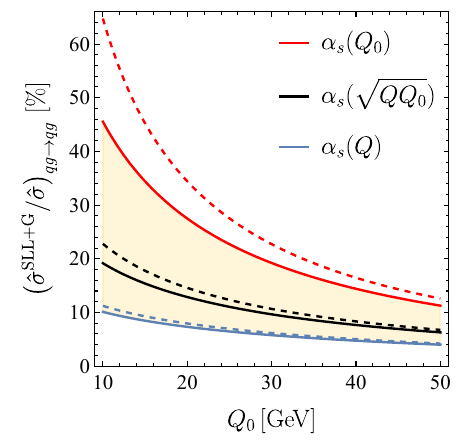}
    \includegraphics[scale=1]{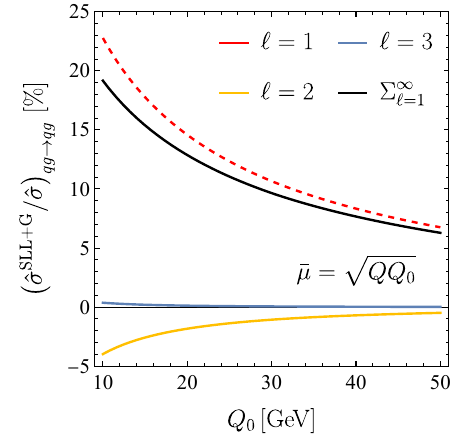}
    \caption{Numerical estimates of the impact of the Glauber series on $qg\to qg$ forward scattering as a function of the jet-veto scale $Q_0$. All plots and curves have the same meaning as in Figure~\ref{fig_gg_2}.}
    \label{fig_qg_qg}
\end{figure}

\begin{figure}[t]
    \centering
    \includegraphics[scale=1]{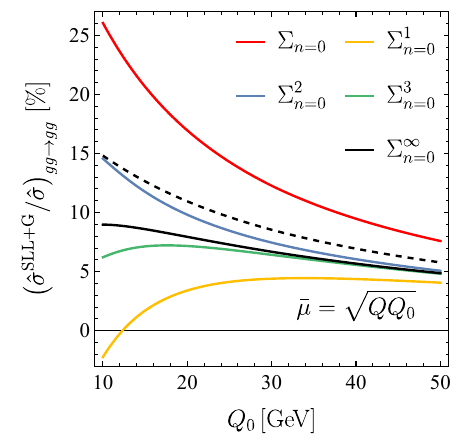}
    \includegraphics[scale=1]{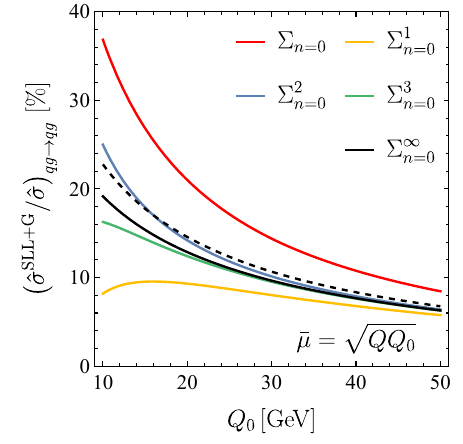}
    \caption{Comparison of the analytic expressions for the resummation of all Glauber phases and up to $n=3$ insertions of $\Gammac$ to the all-order result (black). The plot on the left shows $gg\to gg$ scattering and the one on the right depicts $qg\to qg$ scattering. In both plots $Q=1$\,TeV and $\Delta Y=2$. The dashed curve shows the SLL result.}
    \label{fig_resum_VG_analytic}
\end{figure}

Figures~\ref{fig_gg_2} and~\ref{fig_qg_qg} show the contributions of the Glauber series for these processes as a function of the jet-veto scale $Q_0$.
In the left panels the perturbative uncertainty is estimated by varying the scale $\bar{\mu}$ in the running coupling between the high scale $Q$ and the low scale $Q_0$.
The solid curves show the results for the entire Glauber series, while the dashed ones correspond to the contributions of the SLLs only.
The right panels show the contributions of individual Glauber-phase pairs.
While in all cases the Glauber series is dominated by the two-Glauber contributions (the SLLs), it turns out that in processes with at least one gluon in the initial state, the four-Glauber contribution has a significant effect (with opposite sign).
For the case of $gg \to gg$ scattering this is particularly pronounced.
Yet higher-order Glauber contributions are small in all cases.

\begin{figure}[t]
    \centering
    \includegraphics[scale=1]{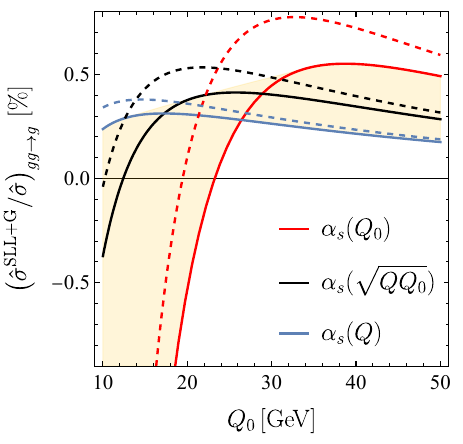}
    \includegraphics[scale=1]{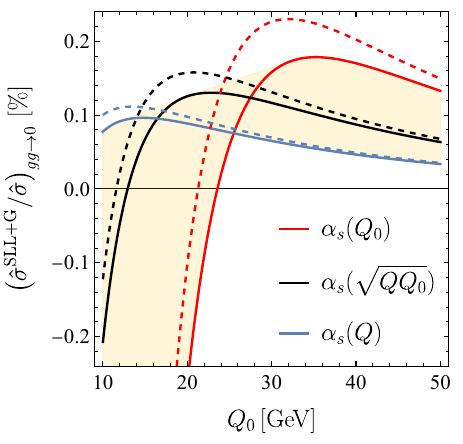}
    \\
    \includegraphics[scale=1]{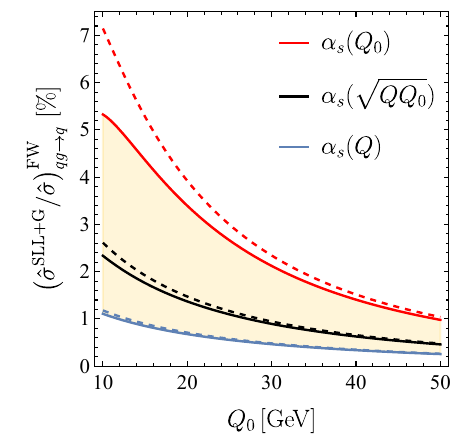}
    \includegraphics[scale=1]{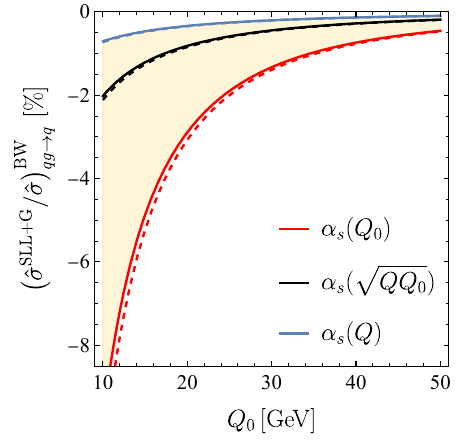}
    \caption{Numerical estimates of the impact of the Glauber series on $gg\to g$ (upper left panel) and $gg\to 0$ (upper right panel) scattering, as well as $qg\to q$ forward (bottom left panel) and backward (bottom right panel) scattering as a function of the jet-veto scale $Q_0$. All figures show the all-order result obtained with three different choices for the scale in the running coupling. Dashed lines show the corresponding curves in the SLL approximation ($\ell=1$). In all plots $Q=1$\,TeV and $\Delta Y=2$.}
    \label{fig_2to1and0}
\end{figure}

Lastly, Figure~\ref{fig_resum_VG_analytic} shows the convergence of the Glauber series when insertions of the Glauber operator $\VG$ are resummed, but $\Gammac$ is treated as a perturbation, for the analytic expressions see~\eqref{eq_resummedVG_0} and~\eqref{eq_resummedVG_1}. 
While the method presented in Section~\ref{sec_resummation} can easily be generalized to larger $n$, i.e.~more insertions of $\Gammac$, the formulas become quite lengthy.
It is noteworthy that already the result for $n=2$ is as good an approximation (or better) to the entire Glauber series than the resummed SLLs. 
For smaller values of $Q_0$, the double-logarithmic corrections dominate the Glauber phases and the approximation becomes worse.

\subsection[Numerical estimates for \texorpdfstring{$2\to 1$}{2->1} and \texorpdfstring{$2\to 0$}{2->0} processes]{Numerical estimates for $\bm{2\to 1}$ and $\bm{2\to 0}$ processes}
\label{subsec_2to1and0}

The leading-order hard functions for $gg\to g$ and $gg\to0$ scattering are given by
\begin{equation}
\begin{split}
    \H_{gg\to g} &= \tr{\H_{gg\to g}} \, \frac{1}{N_c(N_c^2-1)} \, f^{a_1a_2a_3} f^{b_1b_2b_3}\,,
    \\
    \H_{gg\to0} &= \tr{\H_{gg\to0}} \, \frac{1}{N_c^2-1} \, \delta^{a_1a_2} \delta^{b_1b_2} \,.
\end{split}
\end{equation}
In these cases, the color traces with $\Aj$ vanish and only the angular integral $J_{12}$ contributes.
For $qg\to q$ the leading-order hard function reads
\begin{equation}
    \H_{qg\to q} = \tr{\H_{qg\to q}} \, \frac{1}{C_F N_c} \, (t^{a_2})_{\alpha_3\alpha_1} (t^{b_2})_{\beta_1\beta_3} \,,
\end{equation}
which has non-vanishing color traces with $\Oj$ and $\O$.
Therefore, the angular integrals $J_{12}$ and $J_3$ contribute.
In this process forward and backward scattering should be distinguished and one has $J_3=\mp\Delta Y$ in the forward (upper sign) or backward (lower sign) limit~\cite{Becher:2023mtx}.

Figure~\ref{fig_2to1and0} shows the Glauber series contributions for $2\to1$ and $2\to0$ processes, including the uncertainty bands from scale variation.
For $gg\to g$ and $gg\to0$ the Glauber series has a sizeable relative effect compared to the contributions of the SLLs, but the overall magnitude of the effects is very small.
In contrast, for $qg\to q$ scattering one finds larger corrections which, however, are dominated by the two-Glauber contribution. 
It is noteworthy that the effects of the Glauber series in this case differ in sign and shape between forward and backward scattering.

%%%%%%%%%%%%%%%%%%%%%%%%%%%%%%%%%%%%%%%%%%%%%%%%%%%%%%%%%%%%%%%%%%
\section{Conclusions}
\label{sec_conclusions}

In this article, the resummation of the Glauber series for non-global observables at hadron colliders is extended to partonic $2 \to M$ scattering processes featuring gluons in the initial state. 
This series represents the combined sum of large double-logarithms $\ln^2 (Q/Q_0)$ in the ratio of the hard scale $Q$ and the jet-veto scale $Q_0 \ll Q$, together with multiple Glauber-gluon exchanges associated with a large numerical enhancement factor $\pi^2$.
Compared to the scattering of (anti-)quarks studied in~\cite{Boer:2023jsy}, the presence of gluons in the initial state complicates the color algebra considerably.
Nonetheless, the appearing color traces can be reduced in a systematic way by employing a set of tensors in color space that is closed under repeated applications of the Glauber operator $\VG$ and of the soft-collinear, logarithmically-enhanced anomalous dimension $\Gammac$.

The construction of this basis is facilitated by identifying a minimal number of tensors $\F_i$, $\D_i$, $\NABLA_i$ and $\DELTA_i$, see~\eqref{eq_adjoint_matrices}, in the color space of the initial-state particles, which satisfy relations mimicking those of the $SU(N_c)$ color generators.
The basis elements then amount to suitable linear combinations of these tensorial objects and up to one additional final-state generator, see Tables~\ref{tab_Aj_structures}\,--\,\ref{tab_Oj_O_structures}. 
Whereas only four such structures are relevant for quark-initiated scattering processes, the basis consists of thirteen elements for gluon-gluon scattering and eleven elements for quark-gluon scattering. 

The contribution of the Glauber series to a partonic cross section is given by multiple infinite sums, see~\eqref{eq_rlsums}, whose terms are expressed as color traces of these basis structures with process-dependent hard functions.
In the approximation of a scale-independent strong coupling, these terms are given in~\eqref{eq_master_formula}. 
While this multi-sum cannot be cast in a closed analytic form, the series can be truncated at any desired order to obtain numerical results for the cross sections. 

Alternatively, the Glauber series can be organized through a path-ordered evolution operator by restricting the anomalous dimension to only contain $\VG$ and $\Gammac$, see~\eqref{eq_cross_section_Glauber_path_exp}. 
By treating either of these operators perturbatively, one can study the all-order effect of the remaining one.
When expanding in $\Gammac$, this results in multiple insertions of an ordinary exponential of the Glauber operator, while the expansion in $\VG$ gives rise to insertions of a Sudakov-like evolution operator.
The color traces can be computed straightforwardly by employing the basis and only nested scale integrals remain.
If the Glauber phases are resummed, a finite number of $\Gammac$ insertions is included, and the fixed-coupling approximation is used, the series can be expressed in terms of simple trigonometric and polynomial functions in the expansion parameters $w$ and $w_\pi$.
For the case of zero and one $\Gammac$ insertions, explicit results are presented in~\eqref{eq_resummedVG_0} and~\eqref{eq_resummedVG_1}, respectively.

Numerically, higher-order Glauber contributions tend to reduce the effect from the resummation of super-leading logarithms (SLLs), but in most cases their effect is subdominant.
This behavior can be attributed to the fact that the SLLs represent the first term in an alternating series whose coefficients decrease in magnitude. 
Compared with the case of quark-initiated scattering considered in~\cite{Boer:2023jsy}, higher-order Glauber contributions are numerically more important for scattering processes with at least one gluon in the initial state.
Among the channels studied here, the effect of higher Glauber insertions is most pronounced for $gg \to gg$ scattering.
However, for all practical purposes it suffices to consider up to four insertions of the Glauber operator $\VG$.
The perturbative uncertainty from the scale choice in the strong coupling is numerically at least as important as higher Glauber exchanges.
This motivates a study of subleading logarithmic effects in future work. 

Together with~\cite{Becher:2021zkk,Becher:2023mtx,Boer:2023jsy}, this work completes the derivation of analytic expressions of the leading double-logarithmic effects in wide-angle gap-between-jet cross sections for all $2 \to M$ ($M \leq 2$) partonic channels.
A full phenomenological analysis including SLLs and higher-Glauber exchanges beyond the small-angle limit, also including interference effects and parton distribution functions, is left for future work.

\pdfbookmark[1]{Acknowledgements}{Acknowledgements}
\subsubsection*{Acknowledgements}
We are grateful to Thomas Becher and Dominik Schwienbacher for many valuable discussions and comments on the manuscript. 
The authors thank the Albert Einstein Center for Fundamental Physics (AEC) at the University of Bern for hospitality during the final stages of the project.
This work has been supported by the Cluster of Excellence Precision Physics, Fundamental Interactions, and Structure of Matter (PRISMA$^+$ EXC 2118/1) funded by the German Research Foundation (DFG) within the German Excellence Strategy (Project ID 390831469), and has received funding from the European Research Council (ERC) under the European Union’s Horizon 2022 Research and Innovation Programme (Grant agreement No.101097780, EFT4jets).

\clearpage
\begin{appendix}

%%%%%%%%%%%%%%%%%%%%%%%%%%%%%%%%%%%%%%%%%%%%%%%%%%%%%%%%%%%%%%%%%%
\section{Matrix representations of \texorpdfstring{$\VG$}{VG} and \texorpdfstring{$\Gammac$}{Γc}}
\label{app_Matrix_representation}

Below we provide the explicit forms of the matrix representations of the Glauber operator $\VG$ and the collinear-emission operator $\Gammac$ for gluon-gluon and quark-gluon initiated processes. 
The matrices listed here are also given in a supplemented \texttt{Mathematica} notebook.

\subsection{Gluon-initiated processes}
Decomposing $V^G$ for gluon-initiated processes into submatrices as described in~\eqref{eq_VG_decomposition_adjoint} yields
\begin{equation}
	\nu^{(j)} =
	\begin{pmatrix}
		\frac{2}{N_c} & \frac{1}{2} & \frac{1}{N_c} & 0 & 0 & 0 & 0 \\
		0 & -\frac{1}{2} & 0 & 0 & 0 & 0 & 0 \\
		0 & 0 & -1 & \frac{-1}{N_c} & 0 & 0 & \frac{1}{N_c} \\
		0 & 0 & 0 & \frac{-1}{N_c} & 0 & \frac{1}{2} & \frac{-1}{N_c} \\
		\frac{4}{N_c} & 1 & \frac{-2}{N_c} & 0 & 1 & \frac{-2}{N_c} & 1 \\
		0 & 0 & 0 & \frac{1}{N_c} & 0 & \frac{1}{2} & \frac{1}{N_c} \\
		\frac{-8}{N_c} & -3 & 0 & 0 & -1 & \frac{-2}{N_c} & -1
	\end{pmatrix}\,,\quad
	\tilde{\nu}^{(j)} =
	\begin{pmatrix}
		\frac{4}{N_c} & 0 & 0 & 0 & 0 & 0 & 0 \\
		\frac{N_c^2-8}{N_c^2} & -1 & \frac{-1}{N_c} & 0 & 0 & 0 & 0 \\
		\frac{2}{N_c} & 0 & -1 & 0 & \frac{-1}{N_c} & 0 & 0 \\
		1 & -1 & 0 & \frac{-2}{N_c} & 0 & 0 & 0 \\
		-1 & 1 & 0 & 0 & 0 & \frac{-2}{N_c} & 0 \\
		0 & 0 & 0 & 1 & 0 & 1 & 0 \\
		-2 & 0 & \frac{4}{N_c} & 0 & \frac{1}{2} & \frac{2}{N_c} & -\frac{1}{2} \\
	\end{pmatrix} .
\end{equation}
The structure of the matrix $(\Gamma^c)^r$, decomposed as in~\eqref{eq_Gammac_decomposition_adjoint}, is more complicated.
Its submatrices can be expanded in the eleven eigenvalues of $\Gammac$
\begin{equation} \label{eq_eigenvalues_Gammac}
\begin{split}
	v_0=0\,, \qquad v_1=\frac{1}{2}\,, \qquad v_2=1\,, \qquad v_3=\frac{3}{2}\,, \qquad v_4=2 \,, 
	\\
	v_{5,6}=\frac{3N_c\pm2}{2N_c} \,, \qquad v_{7,8}=\frac{2(N_c\pm1)}{N_c} \,, \qquad v_{9,10}= \frac{2N_c\pm1}{N_c} \,,
\end{split}
\end{equation}
where $v_5$, $v_7$ and $v_9$ correspond to the plus signs. 
The eigenvalues $v_{0,1,2,5,6,7,8}$ already appeared in the calculation of the SLL color traces~\cite{Becher:2023mtx}.
For arbitrary $r \geq 0$ the expansion reads
\begin{equation} \label{eq_Gammac_eigenvalue_decomposition}
	\tilde{\gamma}^{(j)}(r) = \sum_{i=0}^{10} v_i^r \; \tilde{\gamma}^{(j)}_i
\end{equation}
and similarly for $\gamma^{(j)}(r)$, $\gamma(r)$ and $\lambda(r)$.
For $v_0$, it is understood that $0^r=\delta_{0r}$.
One obtains
\begin{align}
	\tilde{\gamma}_2^{(j)} &=
	\begin{pmatrix}
		1 & 0 & 0 & 0 & 0 & 0 & 0 \\
		0 & 1 & 0 & 0 & 0 & 0 & 0 \\
		\frac{-4}{N_c} & 0 & 0 & 0 & 0 & 0 & 0 \\
		0 & \frac{2N_c}{N_c^2-4} & 0 & 0 & 0 & 0 & 0 \\
		0 & \frac{2N_c^2}{N_c^2-4} & 0 & 0 & 0 & 0 & 0 \\
		\frac{-2}{N_c} & 0 & 0 & 0 & 0 & 0 & 0 \\
		\frac{2(N_c^2-4)}{-N_c^2} & 0 & 0 & 0 & 0 & 0 & 0
	\end{pmatrix} ,
	&
	\tilde{\gamma}_3^{(j)} &=
	\begin{pmatrix}
		0 & 0 & 0 & 0 & 0 & 0 & 0 \\
		0 & 0 & 0 & 0 & 0 & 0 & 0 \\
		\frac{4}{N_c} & 0 & 1 & 0 & 0 & 0 & 0 \\
		0 & 0 & 0 & 0 & 0 & 0 & 0 \\
		0 & 0 & 0 & 0 & 0 & 0 & 0 \\
		0 & 0 & 0 & 0 & 0 & 0 & 0 \\
		\frac{-16}{N_c^2} & 0 & \frac{-4}{N_c} & 0 & 0 & 0 & 0
	\end{pmatrix} ,
	\nonumber \\
	\tilde{\gamma}_4^{(j)} &=
	\begin{pmatrix}
		0 & 0 & 0 & 0 & 0 & 0 & 0 \\
		0 & 0 & 0 & 0 & 0 & 0 & 0 \\
		0 & 0 & 0 & 0 & 0 & 0 & 0 \\
		0 & 0 & 0 & 0 & 0 & 0 & 0 \\
		0 & 0 & 0 & 0 & 0 & 0 & 0 \\
		\frac{2}{N_c} & 0 & 0 & 0 & 0 & 1 & 0 \\
		\frac{2(N_c^2+4)}{N_c^2} & 0 & \frac{4}{N_c} & 0 & 0 & 0 & 1
	\end{pmatrix} ,
	&
	\tilde{\gamma}_{7,8}^{(j)} &=
	\begin{pmatrix}
		0 & 0 & 0 & 0 & 0 & 0 & 0 \\
		0 & 0 & 0 & 0 & 0 & 0 & 0 \\
		0 & 0 & 0 & 0 & 0 & 0 & 0 \\
		0 & \frac{\pm N_c}{2(N_c\pm2)} & 0 & \frac{1}{2} & \mp\frac{1}{4} & 0 & 0 \\
		0 & \frac{-N_c}{N_c\pm2} & 0 & \mp1 & \frac{1}{2} & 0 & 0 \\
		0 & 0 & 0 & 0 & 0 & 0 & 0 \\
		0 & 0 & 0 & 0 & 0 & 0 & 0
	\end{pmatrix} ,
\end{align}
whereas all other coefficient matrices for $\tilde{\gamma}^{(j)}(r)$ are zero. For $\gamma^{(j)}(r)$ the non-vanishing coefficient matrices are
\begin{align}
	\gamma_1^{(j)} &=
	\begin{pmatrix}
		1 & 0 & 0 & 0 & 0 & 0 & 0 \\
		0 & 0 & 0 & 0 & 0 & 0 & 0 \\
		\frac{2}{N_c^2-1} & 0 & 0 & 0 & 0 & 0 & 0 \\
		\frac{2 N_c}{N_c^2-1} & 0 & 0 & 0 & 0 & 0 & 0 \\
		0 & 0 & 0 & 0 & 0 & 0 & 0 \\
		\frac{-4}{N_c^2-1} & 0 & 0 & 0 & 0 & 0 & 0 \\
		\frac{-4 N_c}{N_c^2-1} & 0 & 0 & 0 & 0 & 0 & 0
	\end{pmatrix} ,
	&
	\gamma_2^{(j)} &=
	\begin{pmatrix}
		0 & 0 & 0 & 0 & 0 & 0 & 0 \\
		0 & 1 & 0 & 0 & 0 & 0 & 0 \\
		0 & \frac{2 N_c}{N_c^2-4} & 0 & 0 & 0 & 0 & 0 \\
		0 & \frac{N_c^2}{N_c^2-4} & 0 & 0 & 0 & 0 & 0 \\
		0 & -1 & 0 & 0 & 0 & 0 & 0 \\
		0 & \frac{-2 N_c}{N_c^2-4} & 0 & 0 & 0 & 0 & 0 \\
		0 & \frac{-N_c^2}{N_c^2-4} & 0 & 0 & 0 & 0 & 0
	\end{pmatrix} ,
	\nonumber \\
	\gamma_3^{(j)} &=
	\begin{pmatrix}
		0 & 0 & 0 & 0 & 0 & 0 & 0 \\
		0 & 0 & 0 & 0 & 0 & 0 & 0 \\
		0 & 0 & 0 & 0 & 0 & 0 & 0 \\
		0 & 0 & 0 & 0 & 0 & 0 & 0 \\
		0 & 1 & 0 & 0 & 1 & 0 & 0 \\
		0 & \frac{-2 N_c}{N_c^2-4} & 0 & 0 & \frac{-2 N_c}{N_c^2-4} & 0 & 0 \\
		0 & \frac{-N_c^2}{N_c^2-4} & 0 & 0 & \frac{-N_c^2}{N_c^2-4} & 0 & 0
	\end{pmatrix} ,
	&
	\gamma_{5,6}^{(j)} &=
	\begin{pmatrix}
		0 & 0 & 0 & 0 & 0 & 0 & 0 \\
		0 & 0 & 0 & 0 & 0 & 0 & 0 \\
		\frac{\pm1}{N_c\pm1} & \frac{\pm N_c}{2 (N_c\pm2)} & \frac{1}{2} & \mp\frac{1}{2} & 0 & 0 & 0 \\
		\frac{-1}{N_c\pm1} & \frac{-N_c}{2 (N_c\pm2)} & \mp\frac{1}{2} & \frac{1}{2} & 0 & 0 & 0 \\
		0 & 0 & 0 & 0 & 0 & 0 & 0 \\
		\frac{2}{N_c (N_c\pm1)} & \frac{1}{N_c\pm2} & \frac{\pm1}{N_c} & \frac{-1}{N_c} & 0 & 0 & 0 \\
		\frac{\mp2}{N_c (N_c\pm1)} & \frac{\mp1}{N_c\pm2} & \frac{-1}{N_c} & \frac{\pm1}{N_c} & 0 & 0 & 0
	\end{pmatrix} ,
	\nonumber \\
	\gamma_{9,10}^{(j)} &=
	\begin{pmatrix}
		0 & 0 & 0 & 0 & 0 & 0 & 0 \\
		0 & 0 & 0 & 0 & 0 & 0 & 0 \\
		0 & 0 & 0 & 0 & 0 & 0 & 0 \\
		0 & 0 & 0 & 0 & 0 & 0 & 0 \\
		0 & 0 & 0 & 0 & 0 & 0 & 0 \\
		\frac{\mp2}{N_c} & \frac{\mp(N_c\pm1)}{N_c\pm2} & \frac{\mp1}{N_c} & \frac{1}{N_c} & \frac{\mp N_c}{2 (N_c\pm2)} & \frac{1}{2} & \mp\frac{1}{2} \\
		\frac{2}{N_c} & \frac{N_c\pm1}{N_c\pm2} & \frac{1}{N_c} & \frac{\mp1}{N_c} & \frac{N_c}{2 (N_c\pm2)} & \mp\frac{1}{2} & \frac{1}{2}
	\end{pmatrix} .
	\hspace{-\textwidth} &&
\end{align}
The non-vanishing coefficient matrices for $\gamma(r)$ are given by
\begin{equation}
\begin{aligned}
	\gamma_0 &=
	\begin{pmatrix}
		1 & 0 & 0 & 0 & 0 & 0 \\
		0 & 0 & 0 & 0 & 0 & 0 \\
		0 & 0 & 0 & 0 & 0 & 0 \\
		\frac{4}{N_c^2-1} & 0 & 0 & 0 & 0 & 0 \\
		\frac{8 N_c}{N_c^2-1} & 0 & 0 & 0 & 0 & 0 \\
		0 & 0 & 0 & 0 & 0 & 0
	\end{pmatrix} ,
	& \qquad
	\gamma_2 &=
	\begin{pmatrix}
		0 & 0 & 0 & 0 & 0 & 0 \\
		0 & 1 & 0 & 0 & 0 & 0 \\
		0 & 0 & 1 & 0 & 0 & 0 \\
		0 & 0 & \frac{4 N_c}{N_c^2-4} & 0 & 0 & 0 \\
		0 & 0 & \frac{4 N_c^2}{N_c^2-4} & 0 & 0 & 0 \\
		0 & \frac{-4}{N_c} & 0 & 0 & 0 & 0
	\end{pmatrix} ,
	\\
	\gamma_4 &=
	\begin{pmatrix}
		0 & 0 & 0 & 0 & 0 & 0 \\
		0 & 0 & 0 & 0 & 0 & 0 \\
		0 & 0 & 0 & 0 & 0 & 0 \\
		0 & 0 & 0 & 0 & 0 & 0 \\
		0 & 0 & 0 & 0 & 0 & 0 \\
		0 & \frac{4}{N_c} & 0 & 0 & 1 & 0
	\end{pmatrix} ,
	& \qquad
	\gamma_{7,8} &=
	\begin{pmatrix}
		0 & 0 & 0 & 0 & 0 & 0 \\
		0 & 0 & 0 & 0 & 0 & 0 \\
		0 & 0 & 0 & 0 & 0 & 0 \\
		\frac{\pm2}{N_c\pm1} & 0 & \frac{\pm N_c}{N_c\pm2} & \frac{1}{2} & \mp\frac{1}{4} & 0 \\
		\frac{-4}{N_c\pm1} & 0 & \frac{-2N_c}{N_c\pm2} & \mp1 & \frac{1}{2} & 0 \\
		0 & 0 & 0 & 0 & 0 & 0
	\end{pmatrix} .
\end{aligned}
\end{equation}
For matrix $\lambda(r)$ all eigenvalues contribute
\begin{align}
	\lambda_0 &=
	\begin{pmatrix}
		2 N_c & 0 & 0 & 0 & 0 & 0 \\
		0 & 0 & 0 & 0 & 0 & 0 \\
		\frac{4 N_c}{N_c^2-1} & 0 & 0 & 0 & 0 & 0 \\
		\frac{4 N_c^2}{N_c^2-1} & 0 & 0 & 0 & 0 & 0 \\
		0 & 0 & 0 & 0 & 0 & 0 \\
		\frac{-8 N_c}{N_c^2-1} & 0 & 0 & 0 & 0 & 0 \\
		\frac{-8 N_c^2}{N_c^2-1} & 0 & 0 & 0 & 0 & 0
	\end{pmatrix} ,
	&
	\lambda_1 &=
	\begin{pmatrix}
		-2 N_c & 2 & 0 & 0 & 0 & 0 \\
		0 & 0 & 0 & 0 & 0 & 0 \\
		\frac{-4 N_c}{N_c^2-1} & \frac{4}{N_c^2-1} & 0 & 0 & 0 & 0 \\
		\frac{-4 N_c^2}{N_c^2-1} & \frac{4 N_c}{N_c^2-1} & 0 & 0 & 0 & 0 \\
		0 & 0 & 0 & 0 & 0 & 0 \\
		\frac{8 N_c}{N_c^2-1} & \frac{-8}{N_c^2-1} & 0 & 0 & 0 & 0 \\
		\frac{8 N_c^2}{N_c^2-1} & \frac{-8 N_c}{N_c^2-1} & 0 & 0 & 0 & 0
	\end{pmatrix} ,
	\nonumber \\
	\lambda_2 &=
	\begin{pmatrix}
		0 & -2 & 0 & 0 & 0 & 0 \\
		0 & 0 & 0 & 0 & 0 & 0 \\
		0 & 4 & 0 & 0 & 0 & 0 \\
		0 & -2 N_c & 0 & 0 & 0 & 0 \\
		0 & 0 & 2 N_c & 0 & 0 & 0 \\
		0 & 4 & \frac{-4 N_c^2}{N_c^2-4} & 0 & 0 & 0 \\
		0 & 0 & \frac{-2 N_c^3}{N_c^2-4} & 0 & 0 & 0
	\end{pmatrix} ,
	&
	\lambda_3 &= 
	\begin{pmatrix}
		0 & 0 & 0 & 0 & 0 & 0 \\
		0 & 0 & 0 & 0 & 0 & 0 \\
		0 & 0 & 0 & 0 & 0 & 0 \\
		0 & 0 & 0 & 0 & 0 & 0 \\
		0 & \frac{-4}{N_c} & -2 N_c & 0 & 0 & -1 \\
		0 & \frac{8}{N_c^2-4} & \frac{4 N_c^2}{N_c^2-4} & 0 & 0 & \frac{2 N_c}{N_c^2-4} \\
		0 & \frac{4 N_c}{N_c^2-4} & \frac{2 N_c^3}{N_c^2-4} & 0 & 0 & \frac{N_c^2}{N_c^2-4}
	\end{pmatrix} ,
	\nonumber \\
	\lambda_4 &=
	\begin{pmatrix}
		0 & 0 & 0 & 0 & 0 & 0 \\
		0 & 0 & 0 & 0 & 0 & 0 \\
		0 & 0 & 0 & 0 & 0 & 0 \\
		0 & 0 & 0 & 0 & 0 & 0 \\
		0 & \frac{4}{N_c} & 0 & 0 & 0 & 1 \\
		0 & -4 & 0 & 0 & 0 & -N_c \\
		0 & 0 & 0 & 0 & 0 & 0
	\end{pmatrix} ,
	&
	\lambda_{5,6} &=
	\begin{pmatrix}
		0 & 0 & 0 & 0 & 0 & 0 \\
		0 & 0 & 0 & 0 & 0 & 0 \\
		\frac{2}{N_c\pm1} & \frac{N_c (N_c\pm3)}{\mp (N_c\pm1)} & \frac{N_c}{N_c\pm2} & \pm\frac{1}{2} & -\frac{1}{4} & 0 \\
		\frac{\mp2}{N_c\pm1} & \frac{N_c (N_c\pm3)}{N_c\pm1} & \frac{\mp N_c}{N_c\pm2} & -\frac{1}{2} & \pm\frac{1}{4} & 0 \\
		0 & 0 & 0 & 0 & 0 & 0 \\
		\frac{\pm4}{N_c (N_c\pm1)} & \frac{2 (N_c\pm3)}{-(N_c\pm1)} & \frac{\pm2}{N_c\pm2} & \frac{1}{N_c} & \frac{\mp1}{2 N_c} & 0 \\
		\frac{-4}{N_c (N_c\pm1)} & \frac{2 (N_c\pm3)}{\pm(N_c\pm1)} & \frac{-2}{N_c\pm2} & \frac{\mp1}{N_c} & \frac{1}{2 N_c} & 0
	\end{pmatrix} ,
	\nonumber \\
	\lambda_{7,8} &=
	\begin{pmatrix}
		0 & 0 & 0 & 0 & 0 & 0 \\
		0 & 0 & 0 & 0 & 0 & 0 \\
		\frac{-2}{N_c\pm1} & 0 & \frac{-N_c}{N_c\pm2} & \mp\frac{1}{2} & \frac{1}{4} & 0 \\
		\frac{\pm2}{N_c\pm1} & 0 & \frac{\pm N_c}{N_c\pm2} & \frac{1}{2} & \mp\frac{1}{4} & 0 \\
		0 & 0 & 0 & 0 & 0 & 0 \\
		\frac{2 (N_c\mp1)}{\mp(N_c\pm1)} & 0 & \frac{N_c (N_c\mp1)}{\mp(N_c\pm2)} & \frac{N_c\mp1}{-2} & \frac{N_c\mp1}{\pm4} & 0 \\
		\frac{2 (N_c\mp1)}{N_c\pm1} & 0 & \frac{N_c (N_c\mp1)}{N_c\pm2} & \frac{N_c\mp1}{\pm2} & \frac{N_c\mp1}{-4} & 0
	\end{pmatrix} ,
	\hspace{-\textwidth} &&
	\nonumber \\
	\lambda_{9,10} &=
	\begin{pmatrix}
		0 & 0 & 0 & 0 & 0 & 0 \\
		0 & 0 & 0 & 0 & 0 & 0 \\
		0 & 0 & 0 & 0 & 0 & 0 \\
		0 & 0 & 0 & 0 & 0 & 0 \\
		0 & 0 & 0 & 0 & 0 & 0 \\
		\frac{2 (N_c\mp2)}{\pm N_c} & \frac{2 (N_c\pm3)}{N_c\pm2} & \frac{(N_c\mp2) (N_c\pm1)}{\pm(N_c\pm2)} & \frac{(N_c\mp2) (N_c\pm1)}{2 N_c} & \frac{(N_c\mp2) (N_c\pm1)}{\mp4 N_c} & \frac{N_c (N_c\pm3)}{2 (N_c\pm2)} \\
		\frac{2 (N_c\mp2)}{-N_c} & \frac{2 (N_c\pm3)}{\mp(N_c\pm2)} & \frac{(N_c\mp2) (N_c\pm1)}{-(N_c\pm2)} & \frac{(N_c\mp2) (N_c\pm1)}{\mp2 N_c} & \frac{(N_c\mp2) (N_c\pm1)}{4 N_c} & \frac{N_c (N_c\pm3)}{\mp2 (N_c\pm2)}
	\end{pmatrix} .
	\hspace{-\textwidth} &&
\end{align}

\subsection{Quark-gluon-initiated processes}
The same decomposition of $V^G$ for quark-gluon-initiated processes into submatrices as described in~\eqref{eq_VG_decomposition_mixed} gives
\begin{equation}
	\nu^{(j)} =
    \begin{pmatrix}
        -\frac{1}{N_c} & \frac{2}{N_c} & -\frac{1}{2} & \frac{1}{2} & \frac{1}{N_c} & 0 & 0 & 0 \\
        0 & 0 & \frac{1}{2} & -\frac{1}{2} & 0 & 0 & -\frac{1}{N_c} & 0 \\
        \frac{2}{N_c^2} & 0 & \frac{2}{N_c} & 0 & -1 & -\frac{1}{N_c} & 0 & -\frac{1}{N_c}
    \end{pmatrix}
\,,\quad
	\tilde{\nu}^{(j)} =
    \begin{pmatrix}
        -\frac{2}{N_c} & 0 & 0 \\
         \frac{2}{N_c} & 0 & 0 \\
         -\frac{N_c^2-4}{2 N_c^2} & \frac{1}{2} & 0 \\
         \frac{N_c^2-8}{2 N_c^2} & -\frac{N_c^2-4}{2 N_c^2} & -\frac{1}{N_c} \\
         \frac{1}{N_c} & -\frac{1}{N_c} & -1 \\
         \frac{1}{2} & -\frac{1}{2} & 0 \\
         0 & -\frac{2}{N_c} & 0 \\
         -\frac{1}{2} & \frac{1}{2} & -\frac{1}{N_c}
    \end{pmatrix} .
\end{equation}
Expanding the submatrices~\eqref{eq_Gammac_decomposition_mixed} of $(\Gamma^c)^r$ in eigenvalues of $\Gammac$, cf.~\eqref{eq_eigenvalues_Gammac}, only $v_{0,1,2,3,5,6}$ contribute.
The only non-vanishing coefficient matrices of $\tilde{\gamma}^{(j)}(r)$ are
\begin{align}
    \tilde{\gamma}_2^{(j)} &=
    \begin{pmatrix}
         1 & 0 & 0 \\
         0 & 1 & 0 \\
         \frac{-2}{N_c} & 0 & 0
    \end{pmatrix} ,
    &
    \tilde{\gamma}_3^{(j)} &=
    \begin{pmatrix}
         0 & 0 & 0 \\
         0 & 0 & 0 \\
         \frac{2}{N_c} & 0 & 1
    \end{pmatrix} .
\end{align}
For $\gamma^{(j)}(r)$ all eigenvalues except $v_0$ contribute and one finds
\begin{equation}
\begin{aligned}
    \gamma_1^{(j)} &=
    \begin{pmatrix}
         1 & 0 & 0 & 0 & 0 & 0 & 0 & 0 \\
         0 & 1 & 0 & 0 & 0 & 0 & 0 & 0 \\
         0 & 0 & 0 & 0 & 0 & 0 & 0 & 0 \\
         0 & 0 & 0 & 0 & 0 & 0 & 0 & 0 \\
         0 & \frac{2}{N_c^2-1} & 0 & 0 & 0 & 0 & 0 & 0 \\
         0 & \frac{2 N_c}{N_c^2-1} & 0 & 0 & 0 & 0 & 0 & 0 \\
         0 & 0 & 0 & 0 & 0 & 0 & 1 & 0 \\
         0 & 0 & 0 & 0 & 0 & 0 & 0 & 0
    \end{pmatrix} ,
    & \qquad
    \gamma_2^{(j)} &=
    \begin{pmatrix}
         0 & 0 & 0 & 0 & 0 & 0 & 0 & 0 \\
         0 & 0 & 0 & 0 & 0 & 0 & 0 & 0 \\
         0 & 0 & 1 & 0 & 0 & 0 & 0 & 0 \\
         0 & 0 & 0 & 1 & 0 & 0 & 0 & 0 \\
         0 & 0 & 0 & \frac{2 N_c}{N_c^2-4} & 0 & 0 & 0 & 0 \\
         0 & 0 & 0 & \frac{N_c^2}{N_c^2-4} & 0 & 0 & 0 & 0 \\
         0 & 0 & 0 & 0 & 0 & 0 & 0 & 0 \\
         0 & 0 & 1 & 0 & 0 & 0 & 0 & 0
    \end{pmatrix} ,
    \\
    \gamma_3^{(j)} &=
    \begin{pmatrix}
         0 & 0 & 0 & 0 & 0 & 0 & 0 & 0 \\
         0 & 0 & 0 & 0 & 0 & 0 & 0 & 0 \\
         0 & 0 & 0 & 0 & 0 & 0 & 0 & 0 \\
         0 & 0 & 0 & 0 & 0 & 0 & 0 & 0 \\
         0 & 0 & 0 & 0 & 0 & 0 & 0 & 0 \\
         0 & 0 & 0 & 0 & 0 & 0 & 0 & 0 \\
         0 & 0 & 0 & 0 & 0 & 0 & 0 & 0 \\
         0 & 0 & -1 & 0 & 0 & 0 & 0 & 1
    \end{pmatrix} ,
    & \qquad
    \gamma_{5,6}^{(j)} &=
    \begin{pmatrix}
         0 & 0 & 0 & 0 & 0 & 0 & 0 & 0 \\
         0 & 0 & 0 & 0 & 0 & 0 & 0 & 0 \\
         0 & 0 & 0 & 0 & 0 & 0 & 0 & 0 \\
         0 & 0 & 0 & 0 & 0 & 0 & 0 & 0 \\
         0 & \frac{\pm1}{N_c\pm1} & 0 & \frac{\pm N_c}{2(N_c\pm2)} & \frac{1}{2} & \mp\frac{1}{2} & 0 & 0 \\
         0 & \frac{-1}{N_c\pm1} & 0 & -\frac{N_c}{2 (N_c\pm2)} & \mp\frac{1}{2} & \frac{1}{2} & 0 & 0 \\
         0 & 0 & 0 & 0 & 0 & 0 & 0 & 0 \\
         0 & 0 & 0 & 0 & 0 & 0 & 0 & 0
    \end{pmatrix} .
\end{aligned}
\end{equation}
For $\gamma(r)$ there are only two contributions
\begin{align}
    \gamma_0 &=
    \begin{pmatrix}
         1 & 0 & 0 \\
         0 & 0 & 0 \\
         0 & 0 & 0
    \end{pmatrix} ,
    &
    \gamma_2 &=
    \begin{pmatrix}
         0 & 0 & 0 \\
         0 & 1 & 0 \\
         0 & 0 & 1
    \end{pmatrix} .
\end{align}
Like for gluon-initiated processes, all eigenvalues contribute to $\lambda(r)$, i.e.
\begin{align}
    \lambda_0 &=
    \begin{pmatrix}
         -N_c & 0 & 0 \\
         C_F & 0 & 0 \\
         0 & 0 & 0 \\
         0 & 0 & 0 \\
         \frac{1}{N_c} & 0 & 0 \\
         1 & 0 & 0 \\
         0 & 0 & 0 \\
         0 & 0 & 0
    \end{pmatrix} ,
    &
    \lambda_1 &=
    \begin{pmatrix}
         N_c & -1 & 0 \\
         -C_F & 1 & 0 \\
         0 & 0 & 0 \\
         0 & 0 & 0 \\
         -\frac{1}{N_c} & \frac{2}{N_c^2-1} & 0 \\
         -1 & \frac{2 N_c}{N_c^2-1} & 0 \\
         0 & 0 & -1 \\
         0 & 0 & 0
    \end{pmatrix} ,
    &
    \lambda_2 &=
    \begin{pmatrix}
         0 & 1 & 0 \\
         0 & -1 & 0 \\
         0 & 0 & 0 \\
         0 & 0 & 0 \\
         0 & 2 & 0 \\
         0 & -N_c & 0 \\
         0 & 0 & 1 \\
         0 & 0 & N_c
    \end{pmatrix} ,
    \nonumber\\
    \lambda_3 &=
    \begin{pmatrix}
         0 & 0 & 0 \\
         0 & 0 & 0 \\
         0 & 0 & 0 \\
         0 & 0 & 0 \\
         0 & 0 & 0 \\
         0 & 0 & 0 \\
         0 & 0 & 0 \\
         0 & 0 & -N_c
    \end{pmatrix} ,
    &
    \lambda_{5,6} &=
    \begin{pmatrix}
         0 & 0 & 0 \\
         0 & 0 & 0 \\
         0 & 0 & 0 \\
         0 & 0 & 0 \\
         0 & \mp\frac{N_c (N_c\pm3)}{2 (N_c\pm1)} & 0 \\
         0 & \frac{N_c (N_c\pm3)}{2 (N_c\pm1)} & 0 \\
         0 & 0 & 0 \\
         0 & 0 & 0
    \end{pmatrix} .
\end{align}

%%%%%%%%%%%%%%%%%%%%%%%%%%%%%%%%%%%%%%%%%%%%%%%%%%%%%%%%%%%%%%%%%%
\section{Resummation of Glauber phases}
\label{app_resummation}

In the approximation of a fixed coupling $\alpha_s(\bar{\mu})$, the exponentials of $\VG$ in~\eqref{eq_path_exp_expansion_Gammac_terms} can be expressed in the color basis $\bm{X}$ as
\begin{equation} \label{eq_exp_VG_decomposition}
\begin{split}
    \exp\Big(\frac{\alpha_s(\bar{\mu})}{4\pi} \,V^G \, \ln\frac{\mu_2}{\mu_1}\Big) &=
    \begin{pmatrix}
        \mathcal{V}_0 & 0 & 0 \\
        0 & \tilde{\mathcal{V}}_0 & 0 \\
        0 & 0 & 1
    \end{pmatrix}
    \\
    & + \sum_k \cos\Big(v_{\pi,k}\,N_c\,\alpha_s(\bar{\mu})\,\ln\frac{\mu_2}{\mu_1}\Big)
    \begin{pmatrix}
        \mathcal{V}_k & 0 & 0 \\
        0 & \tilde{\mathcal{V}}_k & 0 \\
        0 & 0 & 0
    \end{pmatrix}
    \\
    & + i \sum_k \sin\Big(v_{\pi,k}\,N_c\,\alpha_s(\bar{\mu})\,\ln\frac{\mu_2}{\mu_1}\Big) \, \frac{1}{v_{\pi,k}}
    \begin{pmatrix}
        0 & \mathcal{V}_k\,\nu^{(j)} & 0 \\
        \tilde{\nu}^{(j)} \, \mathcal{V}_k & 0 & 0 \\
        0 & 0 & 0
    \end{pmatrix} ,
\end{split}
\end{equation}
with $\mathcal{V}_0 = 1 - \sum_k \mathcal{V}_k$ and $\tilde{\mathcal{V}}_0 = 1 - \sum_k \tilde{\mathcal{V}}_k$, and
   $\tilde{\mathcal{V}}_k = v_{\pi,k}^{-2} \left( \tilde{\nu}^{(j)} \, \mathcal{V}_k \, \nu^{(j)} \right)$.
Using~\eqref{eq_exp_VG_decomposition}, one can separate the scale dependence and color structure and calculate the resummed result for a different number $n$ of $\Gammac$ insertions.
The result for $n=0$ has been given in~\eqref{eq_resummedVG_0}.
For $n=1$, one obtains
\begin{align} \label{eq_resummedVG_1}
	\hat{\sigma}_{2\to M}^{\rm SLL+G}\big|_{n=1} &= \frac{8\alpha_s(\bar{\mu}) L}{\pi N_c} \, w \bigg\{ \sum_{k=1}^4 \bigg( \frac{\sin{v_{\pi,k}\sqrt{ww_{\pi}}}}{v_{\pi,k}^3(ww_{\pi})^{3/2}} - \frac{1}{v_{\pi,k}^2ww_{\pi}} + \frac{1}{6}\bigg)
	\nonumber\\
	&\quad \times\bigg(\sum_{j=3}^{2+M} J_j \sum_{i\in I^{(j)}} \big(c_{\pi,k}^{(1,0,0)}\big)_i \, \tr{\H_{2\to M}\,\Aj_i} + J_{12} \sum_{i\in I} \big(d_{\pi,k}^{(1,0,0)}\big)_i \, \tr{\H_{2\to M}\,\S_i}\bigg)
	\nonumber\\
	&\quad +\sum_{k_1,k_2=1}^4\bigg(2\,\frac{\sin{v_{\pi,k_1}\sqrt{ww_{\pi}}}}{v_{\pi,k_1}^3(ww_{\pi})^{3/2}} \bigg(\frac{v_{\pi,k_1}\,v_{\pi,k_2}}{v_{\pi,k_1}^2-v_{\pi,k_2}^2}\bigg)^{\!2}
	\nonumber\\
	&\quad\qquad + \frac{\sin{v_{\pi,k_2}\sqrt{ww_{\pi}}}}{v_{\pi,k_2}^3(ww_{\pi})^{3/2}} \, \frac{v_{\pi,k_2}^2 \, (v_{\pi,k_1}^2-3v_{\pi,k_2}^2)}{(v_{\pi,k_1}^2-v_{\pi,k_2}^2)^2} - \frac{\cos{v_{\pi,k_2}\sqrt{ww_{\pi}}}}{v_{\pi,k_2}^2 ww_{\pi}} \, \frac{v_{\pi,k_2}^2}{v_{\pi,k_1}^2-v_{\pi,k_2}^2} \bigg)
	\nonumber\\
	&\quad\qquad \times\sum_{j=3}^{2+M} J_j \sum_{i\in I^{(j)}} \big(c_{\pi,k_1,k_2}^{(0,1,0)}\big)_i \, \tr{\H_{2\to M}\,\Aj_i}
	\nonumber\\
	&\quad +\sum_{k_1, k_2=1}^4 \bigg(\frac{\sin{v_{\pi,k_1}\sqrt{ww_{\pi}}}}{v_{\pi,k_1}^3(ww_{\pi})^{3/2}} \, \frac{v_{\pi,k_1}^2 \, (v_{\pi,k_1}^2+v_{\pi,k_2}^2)}{(v_{\pi,k_1}^2-v_{\pi,k_2}^2)^2} - \frac{1}{v_{\pi,k_2}^2 ww_{\pi}}
	\nonumber\\
	&\quad\qquad + 2\,\frac{\sin{v_{\pi,k_2}\sqrt{ww_{\pi}}}}{v_{\pi,k_2}^3(ww_{\pi})^{3/2}} \, \frac{v_{\pi,k_1}^2 \, (v_{\pi,k_1}^2-2v_{\pi,k_2}^2)}{(v_{\pi,k_1}^2-v_{\pi,k_2}^2)^2} - \frac{\cos{v_{\pi,k_2}\sqrt{ww_{\pi}}}}{v_{\pi,k_2}^2 ww_{\pi}} \, \frac{v_{\pi,k_1}^2}{v_{\pi,k_1}^2-v_{\pi,k_2}^2} \bigg)
	\nonumber\\
	&\quad\qquad \times\sum_{j=3}^{2+M} J_j \sum_{i\in I^{(j)}} \big(c_{\pi,k_1,k_2}^{(0,0,1)}\big)_i \, \tr{\H_{2\to M}\,\Aj_i}
	\bigg\} .
\end{align}
Despite their appearance, the terms in the double sum are well defined for $k_1\to k_2$.
For quark-gluon initiated processes one has to replace $\Aj\to\Oj$ as well as $\S\to\O$ and restrict the sums to the corresponding three eigenvalues.
The new coefficients are given by
\begin{equation}
\begin{aligned}
	c_{\pi,k}^{(1,0,0)} &= \frac{1}{v_{\pi,k}^2} \big(\tilde{c}^{(1|0)} \, \mathcal{V}_k \, \nu^{(j)} \, \gamma^{(j)}(1) \, \mathcal{V}_0 \big) ,
	\qquad &
	c_{\pi,k_1,k_2}^{(0,1,0)} &= \frac{1}{v_{\pi,k_2}^2} \big(\tilde{c}^{(1|0)} \, \mathcal{V}_{k_1} \, \tilde{\gamma}^{(j)}(1) \, \mathcal{V}_{k_2} \, \nu^{(j)} \big) ,
	\\
	d_{\pi,k}^{(1,0,0)} &= \frac{1}{v_{\pi,k}^2} \big(\tilde{c}^{(1|0)} \, \mathcal{V}_k \, \nu^{(j)} \, \lambda(1) \big) ,
    \qquad &
	c_{\pi,k_1,k_2}^{(0,0,1)} &= \frac{1}{v_{\pi,k_1}^2} \big(\tilde{c}^{(1|0)} \, \mathcal{V}_{k_1} \, \nu^{(j)} \, \gamma^{(j)}(1) \, \tilde{\mathcal{V}}_{k_2} \big) .
\end{aligned}
\end{equation}

\subsection{Gluon-initiated processes}
For gluon-initiated processes, the submatrices appearing in~\eqref{eq_exp_VG_decomposition} are given by
\begin{align}
	\mathcal{V}_1 &= \frac{1}{N_c^2-4}
	\begin{pmatrix}
		\frac{N_c^2}{8} & \frac{3N_c^2}{-8} & 0 & \frac{3 N_c}{4 (N_c^2-1)} & \frac{N_c^2}{16} & \frac{3 N_c}{4(N_c^2-1)} & \frac{N_c^2}{-16} \\
		\frac{N_c^2-8}{-8} & \frac{3(N_c^2-8)}{8} & 0 & \frac{N_c^2+8}{4 N_c (N_c^2-1)} & \frac{N_c^2-8}{-16} & \frac{N_c^2+8}{4 N_c (N_c^2-1)} & \frac{N_c^2-8}{16} \\
		0 & 0 & 0 & \frac{N_c^2-4}{N_c^2-1} & 0 & \frac{N_c^2-4}{N_c^2-1} & 0 \\
		\frac{N_c}{-2} & \frac{3 N_c}{2} & 0 & \frac{N_c^4-5 N_c^2-2}{2 (N_c^2-1)} & \frac{N_c}{-4} & \frac{N_c^4-5 N_c^2-2}{2 (N_c^2-1)} & \frac{N_c}{4} \\
		\frac{N_c^2}{4} & \frac{3N_c^2}{-4} & 0 & \frac{N_c (4 N_c^2-19)}{-2 (N_c^2-1)} & \frac{N_c^2}{8} & \frac{N_c (4 N_c^2-19)}{-2 (N_c^2-1)} & \frac{N_c^2}{-8} \\
		\frac{N_c}{2} & \frac{3 N_c}{-2} & 0 & \frac{N_c^4-5 N_c^2+10}{2 (N_c^2-1)} & \frac{N_c}{4} & \frac{N_c^4-5 N_c^2+10}{2 (N_c^2-1)} & \frac{N_c}{-4} \\
		\frac{3 N_c^2-8}{-4} & \frac{3(3 N_c^2-8)}{4} & 0 & \frac{4 N_c^4-19 N_c^2+24}{-2 N_c (N_c^2-1)} & \frac{3 N_c^2-8}{-8} & \frac{4 N_c^4-19 N_c^2+24}{-2 N_c (N_c^2-1)} & \frac{3 N_c^2-8}{8}
	\end{pmatrix} ,
	\nonumber\\
	\mathcal{V}_2 &= \frac{1}{N_c^2-4}
	\begin{pmatrix}
		\frac{N_c^2-5}{2} & \frac{N_c^2-1}{2} & \frac{N_c^2-4}{2 N_c} & \frac{-1}{N_c} & -\frac{1}{4} & \frac{-1}{N_c} & \frac{1}{4} \\
		\frac{N_c^2-5}{2} & \frac{N_c^2-1}{2} & \frac{N_c^2-4}{2 N_c} & \frac{-1}{N_c} & -\frac{1}{4} & \frac{-1}{N_c} & \frac{1}{4} \\
		0 & 0 & 0 & 0 & 0 & 0 & 0 \\
		\frac{N_c^2-8}{-2 N_c} & \frac{3 N_c}{-2} & 0 & \frac{N_c^2-2}{2} & \frac{3 N_c^2-8}{4 N_c} & \frac{N_c^2-6}{-2} & \frac{N_c}{-4} \\
		N_c^2-5 & N_c^2-1 & \frac{N_c^2-4}{N_c} & \frac{-2}{N_c} & -\frac{1}{2} & -\frac{2}{N_c} & \frac{1}{2} \\
		\frac{N_c^2-8}{2 N_c} & \frac{3 N_c}{2} & 0 & \frac{N_c^2-2}{-2} & \frac{3 N_c^2-8}{-4 N_c} & \frac{N_c^2-6}{2} & \frac{N_c}{4} \\
		-\left(N_c^2-5\right) & -\left(N_c^2-1\right) & \frac{N_c^2-4}{-N_c} & \frac{2}{N_c} & \frac{1}{2} & \frac{2}{N_c} & -\frac{1}{2}
	\end{pmatrix} ,
	\\
	\mathcal{V}_{3,4} &= \frac{1}{2}
	\begin{pmatrix}
		\frac{3 N_c\pm4}{8 N_c} & -\frac{1}{8} & \frac{N_c\pm2}{\mp4 N_c} & \frac{\mp1}{4 N_c (N_c\pm1)} & \frac{N_c\pm4}{-16 N_c} & \frac{\mp1}{4 N_c (N_c\pm1)} & \frac{1}{16} \\
		\frac{(3 N_c\pm4) (N_c\mp2)}{-8 N_c (N_c\pm2)} & \frac{N_c\mp2}{8 (N_c\pm2)} & \frac{N_c\mp2}{\pm4 N_c} & \frac{\pm(N_c\mp2)}{4 N_c (N_c\pm1) (N_c\pm2)} & \frac{(N_c\pm4) (N_c\mp2)}{16 N_c (N_c\pm2)} & \frac{\pm(N_c\mp2)}{4 N_c (N_c\pm1) (N_c\pm2)} & \frac{-(N_c\mp2)}{16 (N_c\pm2)} \\
		\frac{\mp(3 N_c\pm4)}{2 (N_c\pm2)} & \frac{\pm N_c}{2 (N_c\pm2)} & 1 & \frac{1}{(N_c\pm1) (N_c\pm2)} & \frac{\pm(N_c\pm4)}{4 (N_c\pm2)} & \frac{1}{(N_c\pm1) (N_c\pm2)} & \frac{\mp N_c}{4(N_c\pm2)} \\
		\frac{3 N_c\pm4}{2 N_c (N_c\pm2)} & \frac{-1}{2 (N_c\pm2)} & \frac{\mp1}{N_c} & \frac{\mp1}{N_c (N_c\pm1) (N_c\pm2)} & \frac{-(N_c\pm4)}{4 N_c (N_c\pm2)} & \frac{\mp1}{N_c (N_c\pm1) (N_c\pm2)} & \frac{1}{4 (N_c\pm2)} \\
		\frac{(3 N_c\mp2) (3 N_c\pm4)}{-4 N_c (N_c\pm2)} & \frac{3 N_c\mp2}{4 (N_c\pm2)} & \frac{3 N_c\mp2}{\pm2 N_c} & \frac{\pm(3 N_c\mp2)}{2 N_c (N_c\pm1) (N_c\pm2)} & \frac{(N_c\pm4) (3 N_c\mp2)}{8 N_c (N_c\pm2)} & \frac{\pm(3 N_c\mp2)}{2 N_c (N_c\pm1) (N_c\pm2)} & \frac{-(3 N_c\mp2)}{8 (N_c\pm2)} \\
		\frac{-(3 N_c\pm4)}{2 N_c (N_c\pm2)} & \frac{1}{2 (N_c\pm2)} & \frac{\pm1}{N_c} & \frac{\pm1}{N_c (N_c\pm1) (N_c\pm2)} & \frac{N_c\pm4}{4 N_c (N_c\pm2)} & \frac{\pm1}{N_c (N_c\pm1) (N_c\pm2)} & \frac{-1}{4 (N_c\pm2)} \\
		\frac{(3 N_c\pm4) (N_c\mp2)}{4 N_c (N_c\pm2)} & \frac{-(N_c\mp2)}{4 (N_c\pm2)} & \frac{N_c\mp2}{\mp2 N_c} & \frac{\mp(N_c\mp2)}{2 N_c (N_c\pm1) (N_c\pm2)} & \frac{(N_c\pm4) (N_c\mp2)}{-8 N_c (N_c\pm2)} & \frac{\mp(N_c\mp2)}{2 N_c (N_c\pm1) (N_c\pm2)} & \frac{N_c\mp2}{8 (N_c\pm2)}\nonumber \\
	\end{pmatrix} .
\end{align}

\subsection{Quark-gluon-initiated processes}
For quark-gluon-initiated processes, the submatrices appearing in~\eqref{eq_exp_VG_decomposition} are given by
\begin{align}
	\mathcal{V}_2 &= \frac{1}{2(N_c^2-1)}
	\begin{pmatrix}
		N_c^2-2 & N_c^2 & N_c \\
		N_c^2-2 & N_c^2 & N_c \\
		0 & 0 & 0
	\end{pmatrix} ,
	&
	\mathcal{V}_{3,4} &= \frac{1}{2}
	\begin{pmatrix}
		\frac{N_c}{2 (N_c\mp1)} & \frac{-N_c}{2 (N_c\mp1)} & \frac{\mp N_c}{2 (N_c\mp1)} \\
		\frac{-(N_c\mp2)}{2 (N_c\mp1)} & \frac{N_c\mp2}{2 (N_c\mp1)} & \frac{\pm(N_c\mp2)}{2 (N_c\mp1)} \\
		\mp1 & \pm1 & 1
	\end{pmatrix} .
\end{align}

\end{appendix}

\clearpage
\pdfbookmark[1]{References}{Refs}
\bibliography{refs.bib}

\end{document}